\documentclass[aps,pre,reprint,superscriptaddress,footnote, twocolumn,notitlepage]{revtex4-1}

\usepackage{tikz}  
\usetikzlibrary{arrows,shapes,positioning,shadows,backgrounds,fit}

\usepackage{relsize}
\usepackage{graphicx}
\usepackage{amsmath, bbm}
\usepackage{amssymb}
\usepackage{amsthm}
\usepackage{mathrsfs}
\usepackage{xcolor}
\usepackage{cancel}
\usepackage{titlesec}
\usepackage{enumitem}  

\usepackage{dsfont}

\newcommand{\eq}[1]{\eqref{#1}}
\newcommand{\fig}[1]{Fig.~\ref{#1}}

\def\be{\begin{eqnarray}}
\def\ee{\end{eqnarray}}
\newcommand{\tr}[1]{\text{Tr}\left(#1\right)}
\newcommand{\<}{\langle}
\renewcommand{\>}{\rangle}
\newcommand{\ket}[1]{|{#1}\rangle}
\newcommand{\bra}[1]{\langle{#1}|}

\newcommand{\avg}[1]{\left\langle {#1} \right\rangle} 
\renewcommand{\ln}[1]{\mathrm{ln} \left({#1} \right)}
\newcommand{\h}{\mathcal{H}}
\newcommand{\bh}{\mathcal{B}(\mathcal{H})}
\newcommand{\trh}{\mathcal{T}(\mathcal{H})}

\newcommand{\ww}{\mathcal{W}}
\newcommand{\uu}{\mathcal{U}}
\newcommand{\pp}{\mathcal{P}}
\newcommand{\loh}{\mathcal{B}(\h)}

\newcommand{\sh}{\mathcal{S}(\h)}

\newcommand{\id}{\mathbb{I}}
\newcommand{\re}{\mathbb{R}}

\newcommand{\lind}{\mathscr{L}}

\newcommand{\g}{{\mathcal{G}}}

\newcommand{\e}{{\mathcal{E}}}
\newcommand{\dd}{{\mathcal{D}}}
\newcommand{\tg}{{\mathcal{E}}}

\theoremstyle{plain}

\newcommand{\hamed}[1]{{ \color{black}{#1}}}

\newcommand{\marti}[1]{{ \color{orange}{#1}}}

\definecolor{myblue}{rgb}{0.2,0.2,0.8}
\definecolor{myblack}{rgb}{0,0,0}
\definecolor{myurl}{rgb}{0.1,0.1,0.4}

\usepackage[colorlinks=true,citecolor=myblue,linkcolor=myblack,urlcolor=myurl]{hyperref}

\begin{document}

\title{
Joint statistics of work and entropy production along quantum trajectories} 

\author{Harry J.~D. Miller}
\affiliation{Department of Physics and Astronomy, The University of Manchester, Manchester M13 9PL, UK.}

\author{M. Hamed Mohammady}
\affiliation{RCQI, Institute of Physics, Slovak Academy of Sciences, D\'ubravsk\'a cesta 9, Bratislava 84511, Slovakia}

\author{Mart\'i Perarnau-Llobet}
    \affiliation{D\' epartement de Physique Appliqu\' ee, Universit\' e de Gen\`eve, Gen\`eve, Switzerland}

\author{Giacomo Guarnieri}
\affiliation{School of Physics, Trinity College Dublin, College Green, Dublin 2, Ireland}
\affiliation{Dahlem Center for Complex Quantum Systems,
Freie Universit\"{a}t Berlin, 14195 Berlin, Germany}


\begin{abstract}
In thermodynamics, entropy production and work quantify irreversibility and the consumption of useful energy, respectively, when a system is driven out of equilibrium. For quantum systems, these quantities can be identified at the stochastic level by unravelling the system's evolution in terms of quantum jump trajectories. We here derive a general formula for computing the joint statistics of work and entropy production in Markovian driven quantum systems, whose instantaneous steady-states are of Gibbs form. If the driven system remains close to the instantaneous Gibbs state at all times, we show that the corresponding two-variable cumulant generating function implies a joint detailed fluctuation theorem so long as detailed balance is satisfied. As a corollary, we derive a modified fluctuation-dissipation relation (FDR) for the entropy production alone, applicable to transitions between arbitrary steady-states, and for systems that violate detailed balance. This FDR contains a term arising from genuinely quantum fluctuations, and extends an analogous relation from classical thermodynamics to the quantum regime.
\end{abstract}

\maketitle

\section{Introduction}

Understanding  the statistics of work, heat, and entropy production is a central aim of the fields of stochastic and quantum thermodynamics \cite{Jarzynski2011,Seifert2012,Esposito2009,campisi2011colloquium,goold2016role}. A natural starting point for studying such fluctuations is to consider a system driven out of thermal equilibrium. One way in which this can be achieved is  by  coupling the system simultaneously to multiple baths at different temperatures and/or chemical potentials. If one only has access to a single bath,  equilibrium can instead be broken by modulating the system’s Hamiltonian in finite time, as well as external parameters such as the temperature of the environment. In this work we focus on the latter situation, where two thermodynamic quantities play a central role: (i) the entropy production, which measures the degree of irreversibility associated with a process, and (ii) the non-adiabatic work, that quantifies the additional contributions to the work done as a system is driven away from its instantaneous Gibbs steady-state. While these quantities are proportional when the system interacts with an environment at a fixed temperature, they become distinct when the environmental temperature changes, and can thus play different roles for describing non-equilibrium behaviour. For microscopic systems, both classical and quantum, non-adiabatic work $\tilde{w}$ and entropy production $\sigma$ are stochastic variables described by a joint probability distribution $P(\sigma,\tilde{w})$. While classically this distribution can be defined from the underlying probabilistic trajectories through the system's phase space \cite{Seifert2012}, this description breaks down in the quantum regime. Instead, one can define fluctuating entropy production and work by measuring quantum jump trajectories \cite{Horowitz2012,Horowitz2013b,Leggio2013a,Horowitz2014,Manzano2015,Liu2016a,Manzano2018,Elouard2018a,Mohammady2019d,Menczel}. These trajectories describe the probabilistic transitions between the states of the system as it exchanges heat with the environment. Fluctuations along a trajectory stem from both quantum-coherent and thermal transitions, and these may be monitored via an external quantum detector \cite{Pekola2013b,Naghiloo2019}.

Studying the joint statistics of stochastic variables such as work and entropy production can provide a more complete description of a thermodynamic process beyond simply focusing on the marginals of $P(\sigma,\tilde{w})$ \cite{Talkner2009,Sinitsyn2011,Campisi2450,Garc}. For example, joint statistics can be used to ascertain general properties surrounding the trade-offs between dissipation and the signal-to-noise ratio of currents in non-equilibrium processes \cite{Guarnieri2019,VuHasegawa2019,Campisi2020}. Understanding such trade-offs can lend insight into the balance between efficiency, power and reliability of microscopic thermal machines \cite{Pietzonka2018,Funo2018a,Holubec2018,Miller2020b}. In this paper we provide a complete characterisation of the distribution $P(\sigma,\tilde{w})$ for Markovian quantum systems, \hamed{whose instantaneous equilibrium states are of Gibbs form}, driven by changes in both external temperature and Hamiltonian parameters. We then focus on the slow driving regime, whereby the system remains close to \hamed{the instantaneous Gibbs state} throughout a process, and derive a general formula that can be used to compute the cumulant generating function for the joint statistics. 
At the technical level, this result can be understood as a linear-response expansion with respect to the driving speed  of all cumulants of work and entropy production  (applying the standard Kubo linear-response formula only gives access to average quantities \cite{Campisi2012geometric,bonancca2014optimal,Ludovico2016adiabatic}), and extends  a similar result obtained in \cite{Scandi2020} for the work statistics at a fixed temperature, which is now obtained within the quantum trajectory approach. Furthermore, this can also be used to recover a number of recent results concerning Landauer erasure \cite{Miller2020} and fluctuation-dissipation relations \cite{Miller2019} in slowly driven systems. 
Going beyond this, \hamed{ we show that in the slow-driving  regime, so long as the detailed balance condition holds, the non-adiabatic wok and entropy production obey  a detailed fluctuation theorem (DFT)} 
\begin{align}\label{eq:DFT}
    \frac{P(\sigma,\tilde{w})}{P(-\sigma,-\tilde{w})}=e^{\sigma}.
\end{align}
This DFT is stronger than the usual joint fluctuation theorem for currents, which only constrains the joint statistics of thermodynamic variables in relation to a hypothetical time-reversed process (see eg. \cite{Andrieux,Garc,Campisi2450}). While the latter relations apply arbitrarily far from equilibrium, we establish the stronger DFT~\eqref{eq:DFT} by restricting our attention to the slow-driving regime. Here the DFT places restrictions on the relationship between the first and second moments of $\tilde{w}$ and $\sigma$. To highlight this we show that~\eqref{eq:DFT} can be used to recover a thermodynamic uncertainty relation (TUR) recently derived in \cite{Miller2020a}. \hamed{Furthermore, when the entropy production alone is considered, we show that systems with arbitrary instantaneous equilibrium states, and which may violate detailed balance,  obey  a quantum generalisation of a classical fluctuation-dissipation relation (FDR)  derived by Mandal and Jarzysnki \cite{Mandal2016a}.} This generalises the quantum FDR derived in \cite{Miller2019,Scandi2020} to Markovian systems that do not need to satisfy quantum detailed balance.    

The structure of the paper is as follows. In Section~\ref{sec:2} we provide an overview of the quantum trajectory approach to thermodynamics. In Section~\ref{sec:3} we determine the non-adiabatic work and entropy production along jump trajectories and present a general expression for the joint cumulant generating function. In Section~\ref{sec:4} we show that this expression takes on a simpler form in the slow driving limit, and provide explicit expressions for the first and second cumulants. In Section~\ref{sec:5} we show that this generating function implies the DFT~\eqref{eq:DFT}, and use this to derive a TUR relating the averages and variances in work to the entropy production. In Section~\ref{sec:6} we derive the FDR for the non-adiabatic entropy production. Finally, in Section~\ref{sec:7} we evaluate the joint statistics for a single ion heat engine in the slow driving limit and numerically verify these inequalities.

\section{Quantum trajectories and the non-adiabatic entropy production for Quantum Markov Semigroups}\label{sec:2}

We will first introduce the definition of quantum jump trajectories for arbitrary open quantum systems undergoing Markovian evolution. Such trajectories describe the dissipative evolution of a driven system induced by interactions with an environment and measurement apparatus. As we will see, each trajectory comes with an associated time-reversed sequence, and interactions with the environment typically break the time-reversal symmetry between forward and reverse paths. Naturally this gives rise to a notion of entropy production, which will measure this degree of time-reversal asymmetry \cite{Esposito2010b,Esposito2018}. We will then demonstrate how this statistical notion of entropy production can be connected with the thermodynamic variables of the system. Our formalism follows closely the approaches taken in \cite{Horowitz2013b,Manzano2015,Manzano2018}.

First, let $\h$ be a complex separable Hilbert space with an algebra of bounded linear operators $\bh$, and the corresponding space of trace-class operators $\trh \subseteq \bh$. Note that for any $A \in \bh$ and $T \in \trh$, we also have $AT, TA \in \trh$. The state space on $\h$ is thus   $\sh := \{\rho \in \trh : \ \rho\geq 0, \  \tr{\rho}= 1 \}$.  Let a family of channels $\e_\lambda := \lbrace \e_\lambda(\theta) : \trh \to \trh, \theta \geq 0 \rbrace$  form a uniformly continuous Quantum Markov Semigroup (QMS) on $\trh$, with bounded generator $\lind_\lambda$.  Here we have parameterised the generator with a finite collection of scalar variables \begin{align}\label{eq:scalars}
     \lambda(t):=\{\lambda^{ext}(t),\Lambda(t) \},
 \end{align}
 to account for any additional time-dependent control during the dynamics, where we assume a slow enough variation of $\lambda(t)$ so that the generator forms a QMS at all times \cite{Albash2012}. The variables $\Lambda$ represent mechanical parameters of the system, while $\lambda^{ext}$ are parameters for the external environment (eg. temperature, chemical potential etc.).  The dual of $\lind_\lambda$, denoted $\lind_\lambda^*$, is defined by the identity  $\tr{\lind_\lambda^*(A) B} = \tr{A \,  \lind_\lambda(B)}$ for all $A\in \bh$ and $B \in \trh$. $\lind_\lambda^*$ generates a unital QMS on $\loh$, i.e. in the Heisenberg picture,   $\e_\lambda^* := \{\e_\lambda^*(\theta) : \loh \to \loh, \theta \geq 0\}$. 
 
 We will assume the existence of a unique faithful steady state $\pi_\lambda$ such that
\begin{align}\label{eq:steady}
    \lim_{\theta\to\infty}e^{\theta \lind_\lambda}(\rho)=\pi_\lambda, \ \ \ \forall \rho\in\sh.
\end{align}
This condition is satisfied if and only if the generator has a non-degenerate zero eigenvalue, and all other eigenvalues have a strictly negative real part (see Theorem 5.4 in \cite{Rivas2011}). We shall denote the steady state in the spectral form $\pi_\lambda = \sum_\mu p_\mu(\lambda) \Pi_\mu(\lambda)$, where $\{p_\mu(\lambda)\}$ is a probability distribution, and $\Pi_\mu(\lambda) = |\pi_\mu(\lambda)\>\<\pi_\mu(\lambda)|$ are rank-1 projections on the eigenstates of $\pi_\lambda$. Since $p_\mu(\lambda) >0$ for all $\mu$,  the steady state is invertible and, as such, we may introduce the so-called \textit{non-equilibrium potential} \cite{Hatano2001b} defined as \begin{align}
   \Phi_\lambda := -\ln{ \pi_\lambda}=\sum_\mu \phi_\mu(\lambda) \Pi_\mu(\lambda),
\end{align}
where  $\phi_\mu(\lambda) := - \ln{p_\mu(\lambda)}$. The current operator associated with this potential is defined by
\begin{align}
    \dot{\Phi}_\lambda:=\frac{d}{dt}\Phi_\lambda.
\end{align}
It is important to stress that in general the potential and current are non-commuting $[\Phi_\lambda, \dot{\Phi}_\lambda]\neq 0$ at any given time. 

As the system's evolution is governed by a QMS, it follows that  the channel describing the system's time  evolution $t = t_1 \mapsto t = t_2 > t_1$ is given by
\begin{align}
\overleftarrow{P}^{t_2}_{t_1} := \overleftarrow{\exp} \bigg(\int^{t_2}_{t_1} dt \ \lind_{\lambda(t)} \bigg).
\end{align}
Consequently, denoting the system's state at time $t$ as $\rho_t$, we have $\rho_{t_2} = \overleftarrow{P}^{t_2}_{t_1} (\rho_{t_1})$.  By making use of the time-splitting formula (Theorem 2.8 \cite{Rivas2011}) we may express the channel $\overleftarrow{P}^{\tau}_{0}$, for any $\tau >0$, in terms of the  limit
\begin{align}\label{eq:split}
  \overleftarrow{P}^{\tau}_{0} =\lim_{\delta t\to 0}\prod^0_{n=N}e^{\delta t\lind_{\lambda_n}}, 
\end{align}
for the sequence $\tau=t_{N+1}\geq ...\geq t_0=0$, where $\delta t= \max|t_{n+1}-t_n|$ and $\lambda_n=\lambda(t_n)$. As such, we denote by $\tg_{n}:= e^{\delta t \lind_{\lambda_n}}$ the instantaneous QMS channels acting on the system at times $t_n$, which we in turn ``unravell'' into the set of  operations  $\tg_{x_n}(\cdot) := K_{x_n}(\lambda_n) (\cdot) K_{x_n}^\dagger(\lambda_n)$,  with the Kraus operators 
\begin{align}\label{eq:kraus}
    &K_{0}(\lambda):=\id-\bigg(iH_{\Lambda}+\frac{1}{2}\sum_x L^\dagger_x(\lambda)L_x(\lambda)\bigg)\delta t, \\
    &K_{x}(\lambda):=L_x(\lambda)\sqrt{\delta t}, \ \ \ \ \ x>0,
\end{align}
so that $\tg_n= \sum_{x_n} \tg_{x_n} $. Here, $H_\Lambda = H_\Lambda^\dagger$ is the Hamiltonian, while $\{L_x(\lambda)\}$ is a collection of ``jump operators'', which by the Gorini-Kossakowski-Sudarshan-Lindblad theorem \cite{Holevo2001} provide a representation of $\lind_\lambda$ as
\begin{align}\label{eq:lind}
    \mathscr{L}_\lambda(\cdot):= -i[H_\Lambda,(\cdot)]+\sum_x \mathscr{D}\big(L_x(\lambda)\big) (\cdot),
\end{align}
where $\mathscr{D}(L)(\cdot)=L(\cdot)L^\dagger-\frac{1}{2}\{L^\dagger L,(\cdot)\}$.

Throughout this manuscript, we will only consider processes that impose transitions between steady states, so that the boundary conditions become
\begin{align}\label{eq:boundary}
    \rho_0=\pi_{\lambda(0)} \equiv \pi_{\lambda_0} \mapsto \rho_\tau=\pi_{\lambda(\tau)} \equiv \pi_{\lambda_{N+1}}.
\end{align}
Consequently, the system's evolution may be decomposed into an ensemble of quantum trajectories \cite{Horowitz2012,Horowitz2013b,Horowitz2014,Manzano2015,Manzano2018,Elouard2018a} 
\begin{align}\label{eq:trajec}
\gamma := \{(\mu, \nu), (x_0, \dots , x_N)\},
\end{align}
with probabilities 
\begin{align}\label{eq:trajecprob-app}
p(\gamma) := p_{\mu}(\lambda_0) \tr{\Pi_{\nu}(\lambda_{N+1}) \tg_{x_N}\circ \dots \circ \tg_{x_0}( \Pi_{\mu}(\lambda_0))}.
\end{align}
Here, $\mu$  and $\nu$ denote the outcomes of projective  measurements performed on the system with respect to the eigenbasis of the equilibrium state $\pi_{\lambda(0)}$ and $\pi_{\lambda(\tau)}$,  at the start and end of the cycle, respectively, where we note that $\Pi_\mu(\lambda_0) \equiv \Pi_\mu(\lambda(0))$ and $\Pi_\nu(\lambda_{N+1}) \equiv \Pi_\nu(\lambda(\tau))$. Additionally,  $(x_0, \dots, x_N)$ indicates the sequence of quantum operations $\tg_{x_n}$ acting on the system due to the unravelling of the Lindbladian dynamics. 

In order to evaluate the entropy production for the trajectories $\gamma$, we must first employ a notion of time-reversed dynamics which is ``dual-reverse'' to the QMS describing the system's evolution forwards in time. We shall denote the dual-reverse of $\tg_n$ as $\dd_n = \sum_{x_n} \dd_{x_n}$. For classical Markov chains one may introduce a time-reversal operation that reverses any sequence of microstates in the chain while preserving their transition probabilities \cite{Norris1997}. The analogue of this time reversal in quantum mechanics amounts to reversing the chain of interactions between system and environment. Following Crooks \cite{Crooks2018}, in order to implicitly define the time-reversed process we require that for any configuration $\lambda_n$, the probability of obtaining the sequence $(x_n,  y_n)$ from the channel $\tg_n$, given a system that is initially in the invariant state $\pi_{\lambda_n}$, equals the probability of obtaining the sequence $(y_n,  x_n)$ from the dual-reverse of the channel, $\dd_n$, i.e. 
\begin{align}\label{eq:dualdef}
    \tr{\dd_{x_n} \circ \dd_{y_n} (\pi_{\lambda_n})} = \tr{\tg_{y_n} \circ \tg_{x_n} (\pi_{\lambda_n})}.
\end{align}
The individual dual-reversed  operations in~\eqref{eq:dualdef} 
\marti{are} constructed as
\begin{align}\label{eq:dual-reverse}
    \dd_{x}  = \pp_{\lambda}^{(s)}\circ \tg_{x}^*\circ \pp_{\lambda}^{(-s)},
\end{align}
 for some $s \in [0,1]$, where we define the maps $\pp_\lambda^{ (\pm s)}(\cdot) := \pi_{\lambda}^{\pm s} (\cdot) \pi_{\lambda}^{\pm (1-s)}$ \cite{Fagnola2007}. Therefore, for each trajectory $\gamma$ we define 
 the dual-reverse trajectory by the reversed sequence
\begin{align}\label{eq:dual-traj}
    \tilde \gamma := \{(\nu, \mu), (x_N, \dots, x_0)\},
\end{align}
with the  probability
\begin{align}\label{eq:backprob}
    \tilde p(\tilde \gamma) &= p_\nu(\lambda_{N+1})\tr{\Pi_\mu(\lambda_0) \dd_{x_0}\circ \dots \circ \dd_{x_N} (\Pi_\nu(\lambda_{N+1})) }.
\end{align}
  The \textit{non-adiabatic entropy production} can thus be defined as a statistical measure of the distinguishability between the forward and dual-reverse trajectories \cite{Esposito2010b}:
\begin{align}\label{eq:entprod1}
    \sigma(\gamma) &:=  \ln{\frac{p(\gamma)}{\tilde p(\tilde \gamma)}}, 
\end{align}
By normalisation of $\tilde p(\tilde \gamma)$, and Jensen's inequality,  implies the integral fluctuation theorem:
\begin{align}\label{eq:integralFT}
    \big\langle e^{-\sigma} \big\rangle=1, \ \implies \langle \sigma \rangle\geq 0.
\end{align}
It is worth remarking at this stage that in principle this quantity is not physically accessible to a single run of an experiment, since it implicitly depends on the full ensemble of trajectories. However, as we will see below, once further assumptions about the generator~\eqref{eq:lind} are made, the entropy production~\eqref{eq:entprod1} provides a tool for establishing predictions about physical quantities such as heat and work that can be measured along individual trajectories.

Without further assumptions $\dd_x$ are neither guaranteed to be operations, which is necessary to ensure that $\tilde p(\tilde \gamma)$ always forms a valid probability distribution, nor are they guaranteed to be unique for all $s \in [0,1]$.  First, let us note that \eq{eq:dual-reverse} allows us to write $\dd =\sum_x \mathcal{D}_x=  e^{\delta t \tilde{\lind_\lambda}^* } $, where the s-dual generators $\tilde{\lind}_\lambda$ are defined  as the solution to
\begin{align}\label{eq:dual-reverse-gen}
    \tr{\pi_{\lambda}^{1-s} \tilde{\lind}_\lambda(A) \pi_{\lambda}^{s} B } =\tr{\pi_{\lambda}^{1-s} A \pi_{\lambda}^{s} \lind_\lambda^*(B)}
\end{align}
for $s\in [0,1]$ and all $A, B \in \bh$. Consequently, $\dd_x$ will be operations if $\tilde \lind_\lambda$ is a valid QMS.  As shown by Fagnola and Umanita (Theorem 3.1 and Proposition 8.1 in \cite{Fagnola2007}), $\tilde{\lind_\lambda}$ forms a QMS on $\loh$ for any $s\in [0,1]$ if and only if 
\begin{align}\label{eq:mod}
   [\lind_\lambda^*,\Omega^{(-i)}_\lambda]=0,
\end{align}
 where $\Omega^{(x)}_{\lambda}(\cdot)=\pi_\lambda^{ix} (\cdot) \pi_\lambda^{-ix}$  denotes the  modular automorphism on $\bh$ generated by the invariant state $\pi_\lambda$, which in turn implies that $ [\lind_\lambda^*,\Omega^{(x)}_\lambda]=0$ for all $ x\in\mathbb{R}$.  Under this assumption one can show that the s-dual generator in~\eqref{eq:dual-reverse-gen} is in fact the same for any choice $s\in [0,1]$, thereby singling out a unique dual-reverse QMS. Imposing the condition of \textit{quantum detailed balance} \cite{Alicki1976} is sufficient to guarantee \eq{eq:mod}.  Detailed balance ensures that the generator of the QMS is related to its s-dual via 
\begin{align}\label{eq:DB}
    \tilde{\mathscr{L}}_\lambda(\cdot)=\mathscr{L}_\lambda^*(\cdot) - 2 i[H_\Lambda, (\cdot)].
\end{align}

 As shown in  Proposition 4.4 of Ref. \cite{Fagnola2007},  the constraint~\eqref{eq:mod} implies the existence of a set $\{H_\Lambda, \lbrace L_x(\lambda) \rbrace \}$ satisfying the following:
\begin{align}\label{eq:priv1}
    \nonumber&\pi_\lambda L_x(\lambda)\pi_\lambda^{-1}=e^{-\Delta \phi_x(\lambda)}L_x(\lambda), \, \, \, \forall x. \\
    &\big[H_\Lambda,\pi_\lambda\big]=\sum_x \big[ L^\dagger_x(\lambda)L_x(\lambda),\pi_\lambda\big]=0.
\end{align}
Here $\Delta \phi_x(\lambda)=\phi_i(\lambda)-\phi_j(\lambda)$ for all $(i,j)$ such that $\bra{\pi_i(\lambda)} L_x(\lambda)\ket{\pi_j(\lambda)}\neq 0$. It is straightforward to see from the privileged representation~\eqref{eq:priv1} that the Kraus operators acting on the system satisfy
\begin{align}\label{eq:priv2}
    \pi_\lambda K_x(\lambda)\pi_\lambda^{-1}=e^{-\Delta \phi_x(\lambda)}K_x(\lambda), \ \ \ \Delta \phi_0(\lambda)=0.
\end{align}
By expanding in the eigenstates of $\pi_\lambda$, this means that the Kraus operators~\eqref{eq:kraus} acting on the system at any given time take the form \cite{Manzano2015}
\begin{align}\label{eq:manz}
    K_x(\lambda)=\sum_{ij} m^x_{ij}(\lambda)\ket{\pi_i(\lambda)}\bra{\pi_j(\lambda)},
\end{align}
with $m^x_{ij}(\lambda)=0$ if $\phi_i(\lambda)-\phi_j(\lambda)\neq\Delta\phi_x(\lambda)$. In other words, all quantum jumps that take place come with a well-defined change in the non-equilibrium potential $\Delta\phi_x(\lambda)$, caused by transitions between superpositions of eigenstates of the fixed point $\pi_\lambda$. Before proceeding, we will now introduce a useful identity stemming from~\eqref{eq:priv2}. For a Kraus operator $K_x(\lambda)$  belonging to the privileged representation~\eqref{eq:priv2}, one has (see Appendix~\ref{app:1} for proof)
\begin{align}\label{eq:priv_ident}
    \pi_\lambda^u K_x(\lambda)\pi_\lambda^{-u}=e^{-u\Delta \phi_x(\lambda)}K_x(\lambda), \ \ \ \ u\in\mathbb{R}.
\end{align}
As such, recalling that $\e_x^*(\cdot) := K_x^\dagger (\cdot) K_x$, then by \eq{eq:dual-reverse} and \eq{eq:priv_ident} it follows that the dual reverse operations are given by  
\begin{align}\label{eq:kraus_dual}
    \dd_{x} = e^{\Delta \phi_{x}(\lambda)}\tg_{x}^* ,
\end{align}
which is notably independent of $s$. Finally, using equations \eq{eq:trajecprob-app},  \eq{eq:backprob}, and \eq{eq:kraus_dual},  we may  reduce \eq{eq:entprod1} to
\begin{align}\label{eq:entprod2}
    \sigma(\gamma) = \Delta s(\nu,\mu) - \sum_{n=0}^N \Delta \phi_{x_n}(\lambda_n),
\end{align}
where $\Delta s(\nu,\mu)=\ln{p_\mu (\lambda(0))}-\ln{p_\nu(\lambda(\tau))}$. The identification of~\eqref{eq:entprod2} as the non-adiabatic entropy production was previously obtained by Manzano \textit{et al} in \cite{Manzano2015}. This quantity can be understood as a sum of the change in surprisal associated with the system's boundary conditions $(\mu,\nu)$ and the total change in the non-equilibrium potential $\Phi_{\lambda(t)}$ along the sequence $(x_0, \dots , x_N)$.

\section{Exact expression of the moment generating function for entropy production and work}\label{sec:3}

So far we have not placed any assumptions about the form of the steady state $\pi_\lambda$. Henceforth we assume the steady state to be a canonical Gibbs ensemble with Hamiltonian $H_{\Lambda}$ and inverse temperature $\beta=1/ T$ (we set $k_B=1$ throughout), driven by temperature variations and changes in a set of mechanical parameters $\Lambda$:
\begin{align}\label{eq:gibbs}
    \pi_{\lambda}=\frac{e^{-\beta H_\Lambda}}{\mathcal{Z}_\lambda}, \ \ \ \lambda=\{\beta,\Lambda\},
\end{align}
with $\mathcal{Z}_\lambda :=\tr{e^{-\beta H_\Lambda}}$ the partition function. The process is then described by a curve in the parameter space
\begin{align}\label{eq:gibbs-protocol}
\lambda:t\mapsto \lambda(t):=\{\beta(t),\Lambda(t) \}.
\end{align}

We note that since the stationary states $\pi_\lambda$ are of Gibbs form, then so long as $\lind_\lambda$ admits a privileged representation it follows that the system will also obey \emph{time-translation covariance}.  Consider the unitary representation of the time-translation group $U: \mathbb{R} \ni g \mapsto U(g) = e^{i g H_\Lambda}$, generated by the Hamiltonian $H_\Lambda \in \loh$. The unital QMS $\{e^{\theta \lind_\lambda^*} : \loh \to \loh, \theta \geq 0\}$ obeys time-translation covariance if for all $\theta>0$ and $g \in \mathbb{R}$, $e^{\theta \lind_\lambda^*} \circ \uu_g = \uu_g \circ e^{\theta \lind_\lambda^*} $,    where $\uu_g(\cdot) := U(g) (\cdot) U^\dagger(g)$.  Alternatively, this condition can be stated as the commutation relation 
\begin{align}\label{eq:time-covariance}
    [\mathscr{L}_\lambda^*, \mathscr{H}_\Lambda]=0,
\end{align}
where we define the superoperator $\mathscr{H}_\Lambda(\cdot) := i [H_\Lambda, (\cdot)]$. However, since  $\pi_\lambda = e^{-\beta H_\Lambda}/ \mathcal{Z}_\lambda$, then  $\Omega^{(x)}_{\lambda}= \uu_{-\beta x }$, and so \eq{eq:mod} implies time-translation covariance \eq{eq:time-covariance}.

When $\pi_\lambda$ is given by \eq{eq:gibbs},  the non-adiabatic entropy production becomes 
\begin{align}\label{eq:entprod_thermal}
    \sigma(\gamma) = \Delta s(\nu,\mu) - \sum_{n=0}^N \beta_n\Delta e_{x_n}(\Lambda_n),
\end{align}
where $\Delta e_{x_n}(\Lambda_n)$ denotes the heat absorption caused by Kraus operator $K_{x_n}$, defined as the difference between energy eigenvalues of the instantaneous Hamiltonian $H(\Lambda_n)$. While entropy production quantifies the irreversibility of the process,  one may also consider the work done on the system. For a system that remains in instantaneous equilibrium, this is a deterministic quantity given by
\begin{align}\label{eq:adiabatic_w}
 \ww := \int_0^\tau dt \tr{\dot{H}_{\Lambda(t)} \pi_{\lambda(t)}}.
\end{align}
This quantity defines the \textit{adiabatic work} done \cite{Brandner2020} and typically cannot be extracted in finite time. Outside the adiabatic limit, the \emph{total work done} becomes a stochastic quantity dependent on the trajectory $\gamma$, which is given by
\begin{align}\label{eq:diss-app}
    w(\gamma) &:= \Delta U (\gamma) - Q(\gamma), \nonumber \\
    & =  \Delta F+T(0)\ln{p_\mu (\lambda(0))}-T(\tau)\ln{p_\nu(\lambda(\tau))}\notag\\
    & \qquad\qquad\qquad-\sum_{n=0}^N \Delta e_{x_n}(\Lambda_n),
\end{align}
where:  $\Delta U(\gamma) := \tr{ H_{\Lambda(\tau)}\Pi_\nu(\lambda(\tau) )} - \tr{ H_{\Lambda(0)} \Pi_\mu(\lambda(0))}$ is the increase in internal energy along the entire trajectory $\gamma$, while $Q(\gamma) := \sum_{n=0}^N \Delta e_{x_n}(\Lambda_n)$ is the total heat absorbed; and $\Delta F=T(0)\text{ln} \ \mathcal{Z}_{\lambda(0)}-T(\tau)\text{ln} \ \mathcal{Z}_{\lambda(\tau)} $ is the change in equilibrium free energy. This identification follows from the steady  state boundary conditions~\eqref{eq:boundary} and the first law of thermodynamics. Throughout this manuscript we will also be concerned with the \textit{non-adiabatic work}, given by the difference between the total work and the adiabatic work:
\begin{align}\label{eq:tilde-diss-app}
    \tilde w(\gamma) := w(\gamma) - \ww.
\end{align}
We wish to study the higher order moments in entropy production and non-adiabatic work, which can be determined from the two-variable moment generating function (MGF), defined via the Laplace transform of the joint distribution $P(\sigma,\tilde w)$. Formally the joint distribution is constructed from
\begin{align}
    P(\sigma,\tilde w)=\sum_{\{\gamma \}}\delta[\sigma-\sigma(\gamma)] \delta[\tilde w - \tilde w(\gamma)]  p(\gamma).
\end{align}
Then the MGF is
\begin{align}\label{eq:mgf1}
    \mathcal{G}_{\sigma,\tilde w}(u,v):=\sum_{\{\gamma \}}p(\gamma) \ e^{-u \sigma(\gamma)-v \tilde w(\gamma)}  , \ \ \ u,v\in\mathbb{R}.
\end{align}
We stress that while we sum over discrete trajectories such as~\eqref{eq:trajec}, one must subsequently take the continuum limit $\delta t\to 0$. As we show in Appendix~\ref{app:b2}, the MGF can be exactly determined using the privileged representation~\eqref{eq:priv1}:  
\begin{align}\label{eq:MGFexact}
    \mathcal{G}_{\sigma,\tilde w}(u,v)=  \text{Tr}\bigg(\overleftarrow{\exp}\bigg(\int^\tau_0 dt \ \mathscr{L}_{\lambda} + \delta\Upsilon^{(u,v)}_{\lambda}\star\bigg)(\pi_{\lambda(0)})\bigg).
\end{align}
where  $A\star(\cdot):=A(\cdot)+(\cdot)A^\dagger$,  and
\begin{align}
    \Upsilon^{(u,v)}_\lambda:=-\int_0^{(u\beta+v)/2} ds \  e^{-s\tilde{H}_\lambda} \dot{H}_\Lambda e^{s \tilde{H}_\lambda}-\frac{u}{2}\dot{\beta} \  H_\Lambda,
\end{align}
with $\tilde{H}_\lambda:=H_\Lambda-F_\lambda \id$, and  shifted operators with respect to equilibrium denoted $\delta A_\lambda:=A_\lambda-\tr{A_\lambda \pi_\lambda}\id$.

Once the MGF is determined, one may introduce the corresponding cumulant generating function (CGF), given by
\begin{align}\label{eq:CGF1}
    \mathcal{K}_{\sigma,\tilde w}:=\ln{\mathcal{G}_{\sigma,\tilde w}(u,v)}.
\end{align}
This function determines the cumulants of the entropy production and non-adiabatic work from
\begin{align}\label{eq:cumulants}
    \nonumber&\langle \Delta\sigma^k \rangle:=(-1)^k\frac{d^k}{du^k}\mathcal{K}_{\sigma,\tilde w}\bigg|_{u=v=0} \\
    &\langle \Delta\tilde{w}^k \rangle:=(-1)^k\frac{d^k}{dv^k}\mathcal{K}_{\sigma,\tilde w}\bigg|_{u=v=0}
\end{align}
with $\langle \Delta\sigma^1 \rangle=\langle \sigma \rangle$ the average, $\langle \Delta\sigma^2 \rangle=\langle \sigma^2 \rangle-\langle \sigma \rangle^2$ the variance, and so on for the higher cumulants of entropy production. For the non-adiabatic work, we note that while $\langle \Delta\tilde{w}^1 \rangle=\langle w \rangle-\ww$, all higher cumulants $k>1$ are in fact equivalent to the cumulants of the total work done~\eqref{eq:diss-app}, namely $\langle \Delta\tilde{w}^k \rangle=\langle \Delta w^k \rangle$.

\section{Slow driving approximation for the CGF}\label{sec:4}

\

\noindent In general, computing the moment generating function~\eqref{eq:MGFexact} is difficult as it requires solving the time-ordered Lindblad master equation. However, if the speed at which the control parameters $\lambda$ are varied is slow compared to the relaxation timescale of the open system dynamics, we can expect the engine to remain close to the instantaneous steady state $\pi_\lambda$ at all times. In this regime the quantum jump trajectories become almost indistinguishable from their time-reversed counterparts~\eqref{eq:dual-traj}, meaning that average entropy production is small. Previously, these approximations have been evaluated for $\langle \sigma \rangle$ in both classical \cite{Sivak2012a} and quantum systems \cite{Scandi}. Here we will perform an analogous approximation of the full MGF~\eqref{eq:MGFexact} for slow transitions between steady states~\eqref{eq:boundary}.  Given that the protocol's duration is $\tau$, we shall define the speed of the protocol as $\epsilon:= 1/\tau$, so that the slow-driving limit is achieved when $(t^{eq}\epsilon)^2 \ll 1$ with $t^{eq}$ the intrinsic relaxation timescale. Note that this timescale is determined by $t^{eq}=1/\Delta g$, where $\Delta g$ is the   spectral gap of the generator \cite{Znidari2015}. If we order the eigenvalues $\{l_n(\mathscr{L}_\lambda)\}$ of $\mathscr{L}_\lambda$ in terms of their real parts, with $l_0(\mathscr{L}_\lambda)=0$, then the spectral gap is equal to the negative real part of the second largest eigenvalue $\Delta g=-\text{Re}(l_1(\mathscr{L}_\lambda))$. For convenience, we shall work in the re-scaled coordinate $t':= \epsilon t$,  so that  $\tilde{\lambda}(t') := \lambda(t)$ and $\tilde{\rho}_{t'}:=\rho_{t}$. Next we need to utilise the \textit{Drazin inverse} $\mathscr{L}_\lambda^+$ for the generator $\mathscr{L}_\lambda$. This superoperator is defined implicitly as the solution to the following set of equations \cite{Boullion1971}:
\begin{enumerate}[label=(\roman*)]
\item $\tr{\mathscr{L}^+_\lambda(A)}=0$ for all $A\in\trh$.
    \item $\mathscr{L}_\lambda\mathscr{L}_\lambda^+(A)=\mathscr{L}_\lambda^+\mathscr{L}_\lambda(A)=A-\tr{ A}\pi_\lambda$.
    \item $\mathscr{L}^+_\lambda(\pi_\lambda)=0$.
\end{enumerate}
One may show that these conditions yield a unique solution given by the following \cite{Scandi}:
\begin{align}\label{eq:drazin}
    \mathscr{L}_\lambda^+(\cdot):=-\int^\infty_0 d\theta \  e^{\theta\mathscr{L}_\lambda}\big((\cdot)-\pi_\lambda \tr{.}\big).
\end{align}
 By introducing the Drazin inverse, the dynamical equation $\dot{\rho}_t=\mathscr{L}_{\lambda(t)}(\rho_t)$ may be inverted in under these rescaled coordinates to give \cite{Cavina2017}:
\begin{align}\label{eq:slow}
      \tilde{\rho}_{t'}=\pi_{\tilde{\lambda} (t')}+\epsilon \ \mathscr{L}_{\tilde{\lambda}(t')}^+ (\dot{\pi}_{\tilde{\lambda}(t')})+\mathcal{O}(\epsilon^2),
\end{align}
 which holds $\forall t'\in[0,1]$. Note that since it is assumed that the derivative of $\lambda$ vanishes at the initial and final point in time, the system begins and ends in the same equilibrium state. We next introduce the quantum covariance \cite{Petz2002,Scandi2020}, which is a non-commutative generalisation of the classical covariance $\text{cov}(a,b)=\langle a b \rangle-\langle a \rangle \langle b \rangle$, defined as
\begin{align}
    \text{cov}^{(s)}_\lambda(A,B):=\tr{A \ \pi^s_\lambda \  B \ \pi^{1-s}_\lambda}-\tr{A \ \pi_\lambda}\tr{B \ \pi_\lambda},
\end{align}
where $s\in\mathbb{R}$. Using this we present the key technical result of this manuscript, with the proof provided in Appendix \ref{app:b}: if the generator $\lind_\lambda$ obeys detailed balance \eq{eq:DB}, the slow-driving approximation of the  CGF~\eqref{eq:CGF1} when $(t^{eq}\epsilon)^2 \ll 1$ reads
\begin{widetext}
\begin{equation}\label{eq:CGF2}
   \mathcal{K}_{\sigma,\tilde w}(u,v)\simeq  -\int_0^\tau dt \ \bigg(  \beta^2 \bar{C}^{(u+Tv)}_\lambda(\dot{H}_\lambda, \dot{H}_\lambda) +(u-u^2) \dot{\beta}^2 C^{(0)}_\lambda(H_\lambda,H_\lambda)+f_T(u,v)\dot{\beta}\beta \ C^{(0)}_\lambda(\dot{H}_\lambda, H_\lambda)\bigg).
\end{equation}
\end{widetext}
Here we define the correlation function 
\begin{align}
C_\lambda^{(s)}(A,B):=\int^\infty_0 d\theta \ \text{cov}^{(s)}_\lambda\big(A(\theta),B(0)\big),
\end{align}
with $A(\theta) :=e^{\theta \mathscr{L}^*_\lambda}(A)$ an observable evolved in the Heisenberg picture at a fixed control parameter $\lambda$. We have also defined the symmeterised correlation function by
\begin{align}\label{eq:covar}
    \bar{C}_\lambda^{(y)}(A,B) :=\int_0^y ds \int_s^{1-s} ds' \ \mathcal{C}_\lambda^{(s')}(A,B).
\end{align}
Additionally the function $f_T(u,v)$ is given by
\begin{align}
f_T(u,v):=Tv-2u(u+Tv-1).
\end{align}
Our approximation~\eqref{eq:CGF2} characterises all work and entropy production cumulants via~\eqref{eq:cumulants}, which now avoids the cumbersome task of computing the time-ordered exponential in~\eqref{eq:MGFexact}. As an example of its application, we can straightforwardly obtain expressions for the average and variance in entropy production, as well as the work done. To do this, we  note the useful Leibniz rule for differentiating integral functions. This states that, given the function $g(t) := \int_{a(t)}^{b(t)} f(z,t) dz$, then 
\begin{align}
    \frac{d}{dt} g(t) = \int_{a(t)} ^{b(t)} \partial_t f(z,t) dz + f(b(t), t) \frac{db}{dt} - f(a(t),t) \frac{da}{dt}. 
\end{align}
Applying this to~\eqref{eq:CGF2} yields the following expressions for the average entropy production and variance:
\begin{align}\label{eq:ent_cumulants}
    &\langle \sigma \rangle\simeq\int^\tau_0 dt  \int^1_0 ds \ C^{(s)}_\lambda(\dot{\Phi}_\lambda,\dot{\Phi}_\lambda), \nonumber \\
    &\langle \Delta \sigma^2 \rangle\simeq\int^\tau_0 dt \ \bigg(C^{(1)}_\lambda(\dot{\Phi}_\lambda,\dot{\Phi}_\lambda)+C^{(0)}_\lambda(\dot{\Phi}_\lambda,\dot{\Phi}_\lambda)\bigg).
\end{align}
Similarly, for the work done we find 
\begin{align}
    &\langle w \rangle\simeq\ww+\int^\tau_0 dt  \int^1_0 ds \ C^{(s)}_\lambda(\dot{H}_\lambda,\dot{\Phi}_\lambda) , \nonumber \\
    &\langle \Delta w^2 \rangle\simeq\int^\tau_0 dt \ \bigg(C^{(1)}_\lambda(\dot{H}_\lambda,\dot{H}_\lambda)+C^{(0)}_\lambda(\dot{H}_\lambda,\dot{H}_\lambda)\bigg).
\end{align}
These expressions for the average work and variance under slow driving were previously obtained in \cite{Miller2019} for the case of a fixed temperature Markovian master equation.

\section{Joint fluctuation theorem and uncertainty relations for work and entropy production}\label{sec:5}

The structure of~\eqref{eq:CGF2} allows one to obtain some notable results on the property of the joint distribution $P(\sigma,\tilde w)$. As we show in Appendix~\ref{app:c}, the CGF satisfies the following symmetry:
\begin{align}\label{eq:sym}
    \mathcal{K}_{\sigma,\tilde w}(u,v)=\mathcal{K}_{\sigma,\tilde w}(1-u,-v).
\end{align}
Taking the inverse Laplace transform of the equivalent relation $\mathcal{G}_{\sigma,\tilde w}(u,v)=\mathcal{G}_{\sigma,\tilde w}(1-u,-v)$, we obtain a detailed fluctuation relation for the non-adiabatic work and entropy production: 
\begin{align}\label{eq:evan}
    \frac{P(\sigma,\tilde w)}{P(-\sigma,-\tilde w)}= e^{\sigma}.
\end{align}
 This type of relation is often referred to as the Evan-Searles or exchange fluctuation theorem  \cite{E2002}. It can be seen as a stronger version of  the standard  fluctuation theorem, since it only involves the probability distribution $P(\sigma,\tilde w)$ and does not depend on the  distribution of the time-reversed process  $\tilde{P}(\sigma,\tilde w)$ (compare with e.g.~\cite{Andrieux,Garc,Campisi2450}). Typically DFTs of the form \eqref{eq:evan} apply to autonomous systems exchanging conserved quantities \cite{Jarzynski2004b,Jeon2017}, or to systems driven by time-symmetric protocols \cite{Crooks}. Our result thus extends the domain of applicability of~\eqref{eq:evan} to arbitrary protocols in the slow driving regime, which is valid whenever the control variables of the system are varied slowly in comparison to the relaxation timescale of the system.

Going further, the fluctuation relation~\eqref{eq:evan} also implies a \textit{thermodynamic uncertainty relation} (TUR) connecting the average work, its fluctuations and dissipation:
 \begin{align}\label{eq:TUR}
     \frac{\langle \Delta w^2 \rangle \langle \sigma \rangle}{(\langle w \rangle-\ww)^2} \geq 2 .
 \end{align}
 This follows from the results established in \cite{Merhav2010,Guarnieri2019,VuHasegawa2019} and the assumption $\langle \sigma \rangle^2\ll 1$ valid for slow driving. In \cite{Miller2020a} we demonstrate that~\eqref{eq:TUR} may be used to derive a finite time correction to the Carnot bound for periodically driven quantum heat engines, and can in fact be made tighter by taking into account the impact of quantum coherence generated during the process. 
 
 \section{Fluctuation-dissipation relation for entropy production beyond detailed balance}\label{sec:6}

Finally, our results can be used to derive a fluctuation-dissipation relation for the non-adiabatic entropy production alone. In fact, to arrive at this we can drop the assumption of a thermal steady state and instead leave this arbitrary, denoted simply by $\pi_\lambda$ such that  $\mathscr{L}_\lambda(\pi_\lambda) =0$. This could include examples such as the generalised Gibbs ensemble \cite{Perarnau-Llobet2016a} or the squeezed thermal state \cite{Manzano2016}.  Furthermore, we may also drop the assumption of detailed balance~\eqref{eq:DB} and instead only require the existence of a privileged representation for the quantum jump trajectories, which is ensured by imposing~\eqref{eq:mod}. Note however that the existence of a privileged representation requires that the stationary state commutes with the Hamiltonian in~\eqref{eq:lind} \cite{Manzano2015}. In this setting we return to the more abstract notion of stochastic entropy production as a measure of time-reversal asymmetry~\eqref{eq:entprod1}, namely $\sigma(\gamma):=  \ln{p(\gamma)/\tilde p(\tilde \gamma)}$. Following the same steps we took to arrive at~\eqref{eq:MGFexact}, one finds the MGF for the entropy production to be  
\begin{align}\label{eq:MGFexact_ent}
    \mathcal{G}_{\sigma}(u)=\text{Tr}\bigg(\overleftarrow{\exp}\bigg(\int^\tau_0 dt \ \mathscr{L}_{\lambda} + \Upsilon^{(u)}_{\lambda}\star\bigg)(\pi_{\lambda(0)})\bigg),
\end{align}
where 
\begin{align}
    \Upsilon^{(u)}_\lambda:=-\int_0^{u/2} ds \  \pi^s_\lambda \ \dot{\Phi}_\lambda \pi_\lambda^{-s}.
\end{align}
As shown in Appendix \ref{app:E},  the  slow-driving approximation for the corresponding CGF will be
\begin{align}\label{eq:CGF_ent}
    \mathcal{K}_{\sigma}(u)\simeq -\int_0^\tau dt \   \bar{C}^{(u)}_\lambda(\dot{\Phi}_\lambda, \dot{\Phi}_\lambda),
\end{align}
where $\bar{C}$ is the correlation function defined in~\eqref{eq:covar}. This generalises the formula derived in \cite{Scandi2020} to master equations with arbitrary steady states without a requirement of detailed balance or time-translational covariance~\eqref{eq:time-covariance}. Following the same steps as the derivation of~\eqref{eq:sym}, the symmetry $\mathcal{K}_{\sigma}(u)=\mathcal{K}_{\sigma}(1-u)$ implies a detailed fluctuation theorem
\begin{align}\label{eq:entprod-DFT}
    \frac{P(\sigma)}{P(-\sigma)}= e^\sigma,
\end{align}
which  implies an inequality between the average and variance in entropy production \cite{Merhav2010,Guarnieri2019,VuHasegawa2019}:
\begin{align}\label{eq:ent-prod-inequality}
    \langle \Delta\sigma^2 \rangle\geq 2\langle \sigma \rangle.
\end{align}
On further inspection, one finds from the expressions~\eqref{eq:ent_cumulants} that the above inequality may be refined as
\begin{align}\label{eq:FDR}
   \langle \Delta\sigma^2 \rangle = 2 \left( \langle \sigma \rangle + \Delta\mathcal{I}_\sigma\right),
\end{align}
 where we identify the positive quantum correction
\begin{align}\label{eq:skew_cov2}
    \Delta\mathcal{I}_\sigma:= \int_0^\tau dt \int^\infty_0 d\theta \ \mathcal{I}_\lambda(\dot{\Phi}_\lambda(\theta),\dot{\Phi}_\lambda(0))\geq 0.
\end{align}
 Here we introduce the skew covariance $ \mathcal{I}_\lambda(A,B):=-\frac{1}{2}\int^1_0 ds \ \tr{[A,\pi_\lambda^s][B,\pi_\lambda^{1-s}]}$, which is a strictly non-classical measure of covariance between the observables $A,B$ \cite{DenesPetz2009,Hansen2008}. The equality~\eqref{eq:FDR} is a quantum fluctuation-dissipation relation (FDR) for the entropy production of general open systems that admit a privileged representation. This generalises the FDR given in \cite{Miller2019}, which was restricted to systems with thermal fixed points under the condition of detailed balance. The term~\eqref{eq:skew_cov2} quantifies any additional quantum fluctuations in the current operator $\dot{\Phi}_\lambda$ during a slow process. Similar to the findings of \cite{Miller2019}, we can conclude that quantum friction \cite{Feldmann,Plastina} with respect to observable $\dot{\Phi}_\lambda$ increases the overall fluctuations in entropy production relative to the average dissipation. This reaffirms a number of recent results demonstrating that quantum coherence is a detrimental resource to thermodynamic processes in the slow driving or linear response regime \cite{Brandner,Miller2020,Brandner2020,Abiuso2020,Abiuso2020a,Menczel}. This result also extends the FDR derived by Mandal \& Jarzysnki \cite{Mandal2016a} that was applicable to the entropy production for transitions between \textit{classical} non-equilibrium steady states. We may also infer that this quantum friction imparts a non-Gaussian shape in the distribution $P(\sigma)$ \cite{Scandi2020}, which contrasts with the expected Gaussian shape found in the classical stochastic regime \cite{Speck}.

\section{Single ion heat engine}\label{sec:7}

To illustrate our derivations for the cumulant generating function, we consider a model of a single ion in contact with a thermal bath, with the driving protocol $\lambda : t \mapsto \lambda(t) := \{ \beta(t),\omega(t)\}$, with $\omega(t)$ the frequency and $\beta(t) := 1/ T(t)$ the inverse temperature  of the thermal bath. This can be used to build a heat engine, as we discuss in Ref. \cite{Miller2020a}. The engine can be modelled  using a master equation for the damped harmonic oscillator:
\begin{align}\label{eq:ion-lindblad}
    \lind_{\lambda}(\cdot) = - i \omega [a_\omega^{\dagger} a_\omega, (\cdot)]+\Gamma(N_\beta+1) \mathcal{D}_{a_\omega}[\cdot]+\Gamma \mathcal{D}_{a_\omega^{\dagger}}[\cdot],
\end{align}
with $\mathcal{D}_X[\cdot] := X (\cdot) X^{\dagger}- \frac{1}{2}\{X^{\dagger}X,(\cdot)\}$. Here the Hamiltonian is $H_\omega=\omega (a^\dagger_\omega a_\omega+\frac{1}{2})$ with $\omega$ the time-dependent frequency, $a_\omega=\sqrt{\omega/2}(x+i p/\omega)$ is the creation operator with unit mass, $\Gamma$ is the damping rate, and $N_\beta :=1/(e^{\beta \omega}-1)$ is the Bose-Einstein distribution. As a technical remark, we note that  the observables of interest for a harmonic oscillator, such as the Hamiltonian, are unbounded, whereas  our results thus far have been framed in terms of bounded operators. Notwithstanding, the model we consider admits a Master equation obeying detailed balance, and thus falls within the domain of applicability of our main results \cite{Fagnola-harmonic-oscillator}.

\begin{figure*}[htbp!]
\begin{center}
\includegraphics[width=0.9\textwidth]{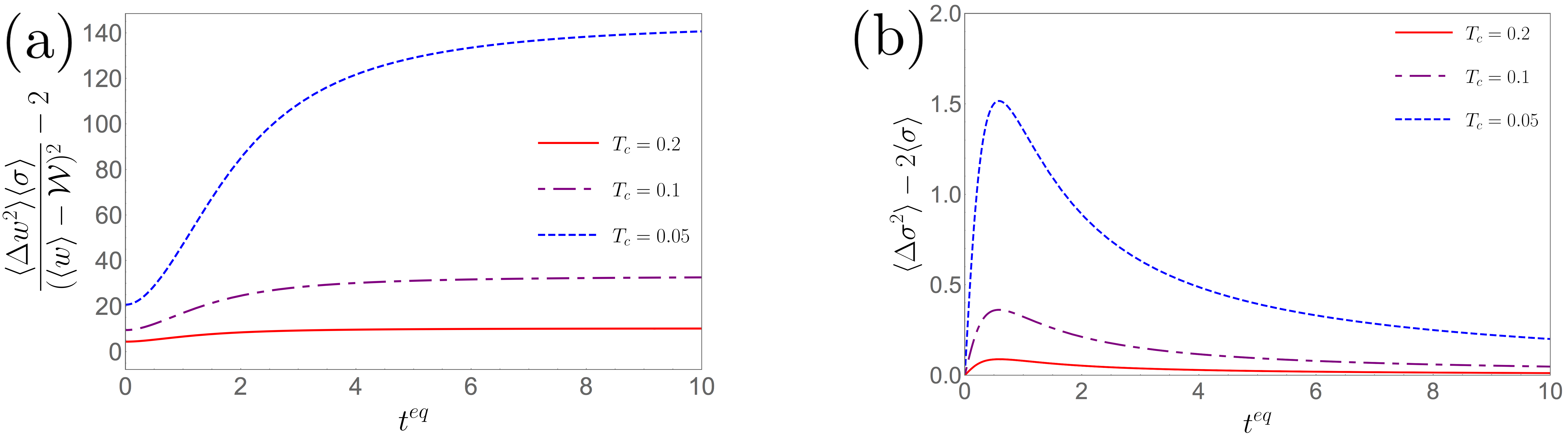}
\vspace*{-0.2cm}
\caption{
Here, we simulate the single ion engine, defined by the protocol in \eq{eq:protocol}, with the parameters $\omega_0=1$,  $T_h=2$, and $\tau = 100$, where we choose units of $\hbar = k_B = 1$. (a)  The TUR  \eq{eq:TUR},  as a function of $t^{eq} :=1/\Gamma$ and  $T_c$.   (b)  The FDR \eq{eq:ent-prod-inequality}  as a function of $t^{eq}:= 1/\Gamma$ and $T_c$. }
\vspace*{-0.5cm}
\label{fig:bound}
\end{center}
\end{figure*}

In Appendix \ref{app:d}, we provide a detailed derivation of  the cumulant generating function $ \mathcal{K}_{\sigma,\tilde w}(u,v)$ for the joint statistics of work and entropy production, given as
\begin{widetext}
\begin{align}
    \mathcal{K}_{\sigma,\tilde w}(u,v) &= -\int_0^\tau dt \ \frac{e^{\beta \omega}}{(e^{\beta \omega}-1)^2} \nonumber \\
    & \qquad \qquad \times \bigg[  \beta^2\dot{\omega}^2   \left(-\frac{\Gamma \sinh{(\beta(u+\beta^{-1} v-1)\omega)} \sinh{(\beta (u+\beta^{-1}v) \omega)}}{\beta^2 \omega^2 (\Gamma^2+4\omega^2)}+\frac{1}{\Gamma}(u+\beta^{-1}v) (1-u-\beta^{-1}v) \right)
    \nonumber\\
    & \qquad \qquad \qquad  +(u-u^2) \dot{\beta}^2\frac{\omega^2}{\gamma}+(\beta^{-1}v-2u(u+\beta^{-1}v-1))\dot{\beta}\beta \ \frac{\omega \dot{\omega}}{\Gamma}\bigg],
\end{align}
\end{widetext}
from which we obtain the first two moments of work and entropy production. These expressions are used in  \cite{Miller2020a} to analyse  the power, efficiency and reliability of a periodically driving heat engine in the slow driving, Markovian regime. In \fig{fig:bound}, we use these  expressions to verify the inequalities \eq{eq:TUR} and \eq{eq:ent-prod-inequality}, for the protocol defined by 
 \begin{align}\label{eq:protocol}
     &\omega(t) = \omega_0 \left(1+\frac{1}{2}\sin\left(\frac{2\pi t}{\tau}\right)+\frac{1}{4}\sin\left(\frac{4\pi t}{\tau} + \pi\right) \right),
     \nonumber\\
     & \beta(t)= \beta_c + (\beta_h - \beta_c)\sin^2 \left(\frac{\pi t}{\tau} \right) ,
\end{align}
where $\beta_c > \beta_h$ and $\omega_0 >0$.

\section{Conclusions}

In this paper we have derived a general expression~\eqref{eq:MGFexact} for the joint cumulant generating function -- of non-adiabatic work and entropy production -- for systems driven away from equilibrium. Our analysis is formulated within the quantum trajectory approach to stochastic thermodynamics and is applicable to Markovian systems governed by a Lindblad master equation, whose instantaneous stationary states are of Gibbs form. Assuming that the Gibbs state is the unique stationary state, and the detailed balance condition holds, we then used adiabatic perturbation theory to derive a simplified expression~\eqref{eq:CGF2} for this function in the regime of slow-driving. From this we were able to obtain a new joint detailed fluctuation theorem for work and entropy production in~\eqref{eq:evan} that holds whenever the system is close to equilibrium throughout the driving. We additionally obtained a slow-driving approximation for the cumulant generating function of entropy production alone \eq{eq:CGF_ent}, valid for arbitrary unique steady states and  systems that do not necessarily fulfill detailed balance, which also lead to a detailed fluctuation theorem \eq{eq:entprod-DFT}.  Finally, we showed that these fluctuation  theorems provide a quantum trajectory derivation of a number of recent results concerning quantum Thermodynamic Uncertainty Relations \cite{Miller2020a} and  Fluctuation Dissipation Relations \cite{Miller2019,Scandi2020}.

\begin{acknowledgments}
H. J. D. M. acknowledges support from the Royal Commission for the Exhibition of 1851. M. H. M. acknowledges support from  the Slovak Academy of Sciences   under MoRePro project OPEQ (19MRP0027), as well as projects OPTIQUTE (APVV-18-0518) and HOQIT (VEGA 2/0161/19).  M. P.-L. acknowledges funding from Swiss National Science Foundation (Ambizione  PZ00P2-186067). G. G. acknowledges fundings from FQXi and DFG Grant No. FOR2724. G. G. also acknowledges funding from European Unions Horizon 2020 research and innovation programme under the Marie Sklodowska-Curie Grant Agreement No. 101026667.
\end{acknowledgments}

\bibliographystyle{apsrev4-1}
\bibliography{mybib}

\begin{thebibliography}{71}%
\makeatletter
\providecommand \@ifxundefined [1]{%
 \@ifx{#1\undefined}
}%
\providecommand \@ifnum [1]{%
 \ifnum #1\expandafter \@firstoftwo
 \else \expandafter \@secondoftwo
 \fi
}%
\providecommand \@ifx [1]{%
 \ifx #1\expandafter \@firstoftwo
 \else \expandafter \@secondoftwo
 \fi
}%
\providecommand \natexlab [1]{#1}%
\providecommand \enquote  [1]{``#1''}%
\providecommand \bibnamefont  [1]{#1}%
\providecommand \bibfnamefont [1]{#1}%
\providecommand \citenamefont [1]{#1}%
\providecommand \href@noop [0]{\@secondoftwo}%
\providecommand \href [0]{\begingroup \@sanitize@url \@href}%
\providecommand \@href[1]{\@@startlink{#1}\@@href}%
\providecommand \@@href[1]{\endgroup#1\@@endlink}%
\providecommand \@sanitize@url [0]{\catcode `\\12\catcode `\$12\catcode
  `\&12\catcode `\#12\catcode `\^12\catcode `\_12\catcode `\%12\relax}%
\providecommand \@@startlink[1]{}%
\providecommand \@@endlink[0]{}%
\providecommand \url  [0]{\begingroup\@sanitize@url \@url }%
\providecommand \@url [1]{\endgroup\@href {#1}{\urlprefix }}%
\providecommand \urlprefix  [0]{URL }%
\providecommand \Eprint [0]{\href }%
\providecommand \doibase [0]{http://dx.doi.org/}%
\providecommand \selectlanguage [0]{\@gobble}%
\providecommand \bibinfo  [0]{\@secondoftwo}%
\providecommand \bibfield  [0]{\@secondoftwo}%
\providecommand \translation [1]{[#1]}%
\providecommand \BibitemOpen [0]{}%
\providecommand \bibitemStop [0]{}%
\providecommand \bibitemNoStop [0]{.\EOS\space}%
\providecommand \EOS [0]{\spacefactor3000\relax}%
\providecommand \BibitemShut  [1]{\csname bibitem#1\endcsname}%
\let\auto@bib@innerbib\@empty
\bibitem [{\citenamefont {Jarzynski}(2011)}]{Jarzynski2011}%
  \BibitemOpen
  \bibfield  {author} {\bibinfo {author} {\bibfnamefont {C.}~\bibnamefont
  {Jarzynski}},\ }\href {\doibase 10.1146/annurev-conmatphys-062910-140506}
  {\bibfield  {journal} {\bibinfo  {journal} {Annual Review of Condensed Matter
  Physics}\ }\textbf {\bibinfo {volume} {2}},\ \bibinfo {pages} {329} (\bibinfo
  {year} {2011})}\BibitemShut {NoStop}%
\bibitem [{\citenamefont {Seifert}(2012)}]{Seifert2012}%
  \BibitemOpen
  \bibfield  {author} {\bibinfo {author} {\bibfnamefont {U.}~\bibnamefont
  {Seifert}},\ }\href {\doibase 10.1088/0034-4885/75/12/126001} {\bibfield
  {journal} {\bibinfo  {journal} {Rep. Prog. Phys}\ }\textbf {\bibinfo {volume}
  {75}},\ \bibinfo {pages} {126001} (\bibinfo {year} {2012})}\BibitemShut
  {NoStop}%
\bibitem [{\citenamefont {Esposito}\ \emph {et~al.}(2009)\citenamefont
  {Esposito}, \citenamefont {Harbola},\ and\ \citenamefont
  {Mukamel}}]{Esposito2009}%
  \BibitemOpen
  \bibfield  {author} {\bibinfo {author} {\bibfnamefont {M.}~\bibnamefont
  {Esposito}}, \bibinfo {author} {\bibfnamefont {U.}~\bibnamefont {Harbola}}, \
  and\ \bibinfo {author} {\bibfnamefont {S.}~\bibnamefont {Mukamel}},\ }\href
  {http://link.aps.org/doi/10.1103/RevModPhys.81.1665} {\bibfield  {journal}
  {\bibinfo  {journal} {Rev. Mod. Phys.}\ }\textbf {\bibinfo {volume} {81}},\
  \bibinfo {pages} {1665} (\bibinfo {year} {2009})}\BibitemShut {NoStop}%
\bibitem [{\citenamefont {Campisi}\ \emph {et~al.}(2011)\citenamefont
  {Campisi}, \citenamefont {H{\"a}nggi},\ and\ \citenamefont
  {Talkner}}]{campisi2011colloquium}%
  \BibitemOpen
  \bibfield  {author} {\bibinfo {author} {\bibfnamefont {M.}~\bibnamefont
  {Campisi}}, \bibinfo {author} {\bibfnamefont {P.}~\bibnamefont {H{\"a}nggi}},
  \ and\ \bibinfo {author} {\bibfnamefont {P.}~\bibnamefont {Talkner}},\ }\href
  {\doibase 10.1103/RevModPhys.83.771} {\bibfield  {journal} {\bibinfo
  {journal} {Rev. Mod. Phys,}\ }\textbf {\bibinfo {volume} {83}},\ \bibinfo
  {pages} {771} (\bibinfo {year} {2011})}\BibitemShut {NoStop}%
\bibitem [{\citenamefont {Goold}\ \emph {et~al.}(2016)\citenamefont {Goold},
  \citenamefont {Huber}, \citenamefont {Riera}, \citenamefont {del Rio},\ and\
  \citenamefont {Skrzypczyk}}]{goold2016role}%
  \BibitemOpen
  \bibfield  {author} {\bibinfo {author} {\bibfnamefont {J.}~\bibnamefont
  {Goold}}, \bibinfo {author} {\bibfnamefont {M.}~\bibnamefont {Huber}},
  \bibinfo {author} {\bibfnamefont {A.}~\bibnamefont {Riera}}, \bibinfo
  {author} {\bibfnamefont {L.}~\bibnamefont {del Rio}}, \ and\ \bibinfo
  {author} {\bibfnamefont {P.}~\bibnamefont {Skrzypczyk}},\ }\href {\doibase
  10.1088/1751-8113/49/14/143001} {\bibfield  {journal} {\bibinfo  {journal}
  {J. Phys. A}\ }\textbf {\bibinfo {volume} {49}},\ \bibinfo {pages} {143001}
  (\bibinfo {year} {2016})}\BibitemShut {NoStop}%
\bibitem [{\citenamefont {Horowitz}(2012)}]{Horowitz2012}%
  \BibitemOpen
  \bibfield  {author} {\bibinfo {author} {\bibfnamefont {J.~M.}\ \bibnamefont
  {Horowitz}},\ }\href {\doibase 10.1103/PhysRevE.85.031110} {\bibfield
  {journal} {\bibinfo  {journal} {Phys. Rev. E}\ }\textbf {\bibinfo {volume}
  {85}},\ \bibinfo {pages} {1} (\bibinfo {year} {2012})}\BibitemShut {NoStop}%
\bibitem [{\citenamefont {Horowitz}\ and\ \citenamefont
  {Parrondo}(2013)}]{Horowitz2013b}%
  \BibitemOpen
  \bibfield  {author} {\bibinfo {author} {\bibfnamefont {J.~M.}\ \bibnamefont
  {Horowitz}}\ and\ \bibinfo {author} {\bibfnamefont {J.~M.~R.}\ \bibnamefont
  {Parrondo}},\ }\href {\doibase 10.1088/1367-2630/15/8/085028} {\bibfield
  {journal} {\bibinfo  {journal} {N. J. Phys}\ }\textbf {\bibinfo {volume}
  {15}},\ \bibinfo {pages} {085028} (\bibinfo {year} {2013})}\BibitemShut
  {NoStop}%
\bibitem [{\citenamefont {Leggio}\ \emph {et~al.}(2013)\citenamefont {Leggio},
  \citenamefont {Napoli}, \citenamefont {Messina},\ and\ \citenamefont
  {Breuer}}]{Leggio2013a}%
  \BibitemOpen
  \bibfield  {author} {\bibinfo {author} {\bibfnamefont {B.}~\bibnamefont
  {Leggio}}, \bibinfo {author} {\bibfnamefont {A.}~\bibnamefont {Napoli}},
  \bibinfo {author} {\bibfnamefont {A.}~\bibnamefont {Messina}}, \ and\
  \bibinfo {author} {\bibfnamefont {H.-P.}\ \bibnamefont {Breuer}},\ }\href
  {\doibase 10.1103/PhysRevA.88.042111} {\bibfield  {journal} {\bibinfo
  {journal} {Phys. Rev. A}\ }\textbf {\bibinfo {volume} {88}},\ \bibinfo
  {pages} {042111} (\bibinfo {year} {2013})}\BibitemShut {NoStop}%
\bibitem [{\citenamefont {Horowitz}\ and\ \citenamefont
  {Sagawa}(2014)}]{Horowitz2014}%
  \BibitemOpen
  \bibfield  {author} {\bibinfo {author} {\bibfnamefont {J.~M.}\ \bibnamefont
  {Horowitz}}\ and\ \bibinfo {author} {\bibfnamefont {T.}~\bibnamefont
  {Sagawa}},\ }\href {\doibase 10.1007/s10955-014-0991-1} {\bibfield  {journal}
  {\bibinfo  {journal} {J. Stat. Phys.}\ }\textbf {\bibinfo {volume} {156}},\
  \bibinfo {pages} {55} (\bibinfo {year} {2014})}\BibitemShut {NoStop}%
\bibitem [{\citenamefont {Manzano}\ \emph {et~al.}(2015)\citenamefont
  {Manzano}, \citenamefont {Horowitz},\ and\ \citenamefont
  {Parrondo}}]{Manzano2015}%
  \BibitemOpen
  \bibfield  {author} {\bibinfo {author} {\bibfnamefont {G.}~\bibnamefont
  {Manzano}}, \bibinfo {author} {\bibfnamefont {J.~M.}\ \bibnamefont
  {Horowitz}}, \ and\ \bibinfo {author} {\bibfnamefont {J.~M.~R.}\ \bibnamefont
  {Parrondo}},\ }\href {\doibase 10.1103/PhysRevE.92.032129} {\bibfield
  {journal} {\bibinfo  {journal} {Phys. Rev. E}\ }\textbf {\bibinfo {volume}
  {92}},\ \bibinfo {pages} {032129} (\bibinfo {year} {2015})}\BibitemShut
  {NoStop}%
\bibitem [{\citenamefont {Liu}(2016)}]{Liu2016a}%
  \BibitemOpen
  \bibfield  {author} {\bibinfo {author} {\bibfnamefont {F.}~\bibnamefont
  {Liu}},\ }\href {\doibase 10.1103/PhysRevE.93.012127} {\bibfield  {journal}
  {\bibinfo  {journal} {Phys. Rev. E}\ }\textbf {\bibinfo {volume} {93}},\
  \bibinfo {pages} {012127} (\bibinfo {year} {2016})}\BibitemShut {NoStop}%
\bibitem [{\citenamefont {Manzano}\ \emph {et~al.}(2018)\citenamefont
  {Manzano}, \citenamefont {Horowitz},\ and\ \citenamefont
  {Parrondo}}]{Manzano2018}%
  \BibitemOpen
  \bibfield  {author} {\bibinfo {author} {\bibfnamefont {G.}~\bibnamefont
  {Manzano}}, \bibinfo {author} {\bibfnamefont {J.~M.}\ \bibnamefont
  {Horowitz}}, \ and\ \bibinfo {author} {\bibfnamefont {J.~M.~R.}\ \bibnamefont
  {Parrondo}},\ }\href {\doibase 10.1103/PhysRevX.8.031037} {\bibfield
  {journal} {\bibinfo  {journal} {Phys. Rev. X}\ }\textbf {\bibinfo {volume}
  {8}},\ \bibinfo {pages} {31037} (\bibinfo {year} {2018})}\BibitemShut
  {NoStop}%
\bibitem [{\citenamefont {Elouard}\ and\ \citenamefont
  {Mohammady}(2018)}]{Elouard2018a}%
  \BibitemOpen
  \bibfield  {author} {\bibinfo {author} {\bibfnamefont {C.}~\bibnamefont
  {Elouard}}\ and\ \bibinfo {author} {\bibfnamefont {M.~H.}\ \bibnamefont
  {Mohammady}},\ }in\ \href {\doibase 10.1007/978-3-319-99046-0_15} {\emph
  {\bibinfo {booktitle} {Thermodynamics in the quantum regime: Fundamental
  Aspects and New Directions}}},\ \bibinfo {series} {Fundamental Theories of
  Physics}, Vol.\ \bibinfo {volume} {195},\ \bibinfo {editor} {edited by\
  \bibinfo {editor} {\bibfnamefont {F.}~\bibnamefont {Binder}}, \bibinfo
  {editor} {\bibfnamefont {L.~A.}\ \bibnamefont {Correa}}, \bibinfo {editor}
  {\bibfnamefont {C.}~\bibnamefont {Gogolin}}, \bibinfo {editor} {\bibfnamefont
  {J.}~\bibnamefont {Anders}}, \ and\ \bibinfo {editor} {\bibfnamefont
  {G.}~\bibnamefont {Adesso}}}\ (\bibinfo  {publisher} {Springer International
  Publishing},\ \bibinfo {address} {Cham},\ \bibinfo {year} {2018})\ pp.\
  \bibinfo {pages} {363--393}\BibitemShut {NoStop}%
\bibitem [{\citenamefont {Mohammady}\ \emph {et~al.}(2020)\citenamefont
  {Mohammady}, \citenamefont {Auff{\`{e}}ves},\ and\ \citenamefont
  {Anders}}]{Mohammady2019d}%
  \BibitemOpen
  \bibfield  {author} {\bibinfo {author} {\bibfnamefont {M.~H.}\ \bibnamefont
  {Mohammady}}, \bibinfo {author} {\bibfnamefont {A.}~\bibnamefont
  {Auff{\`{e}}ves}}, \ and\ \bibinfo {author} {\bibfnamefont {J.}~\bibnamefont
  {Anders}},\ }\href {\doibase 10.1038/s42005-020-0356-9} {\bibfield  {journal}
  {\bibinfo  {journal} {Commun. Phys.}\ }\textbf {\bibinfo {volume} {3}},\
  \bibinfo {pages} {89} (\bibinfo {year} {2020})}\BibitemShut {NoStop}%
\bibitem [{\citenamefont {Menczel}\ \emph {et~al.}(2020)\citenamefont
  {Menczel}, \citenamefont {Flindt},\ and\ \citenamefont {Brandner}}]{Menczel}%
  \BibitemOpen
  \bibfield  {author} {\bibinfo {author} {\bibfnamefont {P.}~\bibnamefont
  {Menczel}}, \bibinfo {author} {\bibfnamefont {C.}~\bibnamefont {Flindt}}, \
  and\ \bibinfo {author} {\bibfnamefont {K.}~\bibnamefont {Brandner}},\
  }\href@noop {} {\bibfield  {journal} {\bibinfo  {journal} {Phys. Rev.
  Research}\ }\textbf {\bibinfo {volume} {2}},\ \bibinfo {pages} {033449}
  (\bibinfo {year} {2020})}\BibitemShut {NoStop}%
\bibitem [{\citenamefont {Pekola}\ \emph {et~al.}(2013)\citenamefont {Pekola},
  \citenamefont {Solinas}, \citenamefont {Shnirman},\ and\ \citenamefont
  {Averin}}]{Pekola2013b}%
  \BibitemOpen
  \bibfield  {author} {\bibinfo {author} {\bibfnamefont {J.~P.}\ \bibnamefont
  {Pekola}}, \bibinfo {author} {\bibfnamefont {P.}~\bibnamefont {Solinas}},
  \bibinfo {author} {\bibfnamefont {A.}~\bibnamefont {Shnirman}}, \ and\
  \bibinfo {author} {\bibfnamefont {D.~V.}\ \bibnamefont {Averin}},\ }\href
  {\doibase 10.1088/1367-2630/15/11/115006} {\bibfield  {journal} {\bibinfo
  {journal} {N. J. Phys}\ }\textbf {\bibinfo {volume} {15}},\ \bibinfo {pages}
  {115006} (\bibinfo {year} {2013})}\BibitemShut {NoStop}%
\bibitem [{\citenamefont {Naghiloo}\ \emph {et~al.}(2020)\citenamefont
  {Naghiloo}, \citenamefont {Tan}, \citenamefont {Harrington}, \citenamefont
  {Alonso}, \citenamefont {Lutz}, \citenamefont {Romito},\ and\ \citenamefont
  {Murch}}]{Naghiloo2019}%
  \BibitemOpen
  \bibfield  {author} {\bibinfo {author} {\bibfnamefont {M.}~\bibnamefont
  {Naghiloo}}, \bibinfo {author} {\bibfnamefont {D.}~\bibnamefont {Tan}},
  \bibinfo {author} {\bibfnamefont {P.~M.}\ \bibnamefont {Harrington}},
  \bibinfo {author} {\bibfnamefont {J.~J.}\ \bibnamefont {Alonso}}, \bibinfo
  {author} {\bibfnamefont {E.}~\bibnamefont {Lutz}}, \bibinfo {author}
  {\bibfnamefont {A.}~\bibnamefont {Romito}}, \ and\ \bibinfo {author}
  {\bibfnamefont {K.~W.}\ \bibnamefont {Murch}},\ }\href {\doibase
  10.1103/PhysRevLett.124.110604} {\bibfield  {journal} {\bibinfo  {journal}
  {Phys. Rev. Lett.}\ }\textbf {\bibinfo {volume} {124}},\ \bibinfo {pages}
  {110604} (\bibinfo {year} {2020})}\BibitemShut {NoStop}%
\bibitem [{\citenamefont {Talkner}\ \emph {et~al.}(2009)\citenamefont
  {Talkner}, \citenamefont {Campisi},\ and\ \citenamefont
  {H{\"{a}}nggi}}]{Talkner2009}%
  \BibitemOpen
  \bibfield  {author} {\bibinfo {author} {\bibfnamefont {P.}~\bibnamefont
  {Talkner}}, \bibinfo {author} {\bibfnamefont {M.}~\bibnamefont {Campisi}}, \
  and\ \bibinfo {author} {\bibfnamefont {P.}~\bibnamefont {H{\"{a}}nggi}},\
  }\href {\doibase 10.1088/1742-5468/2009/02/P02025} {\bibfield  {journal}
  {\bibinfo  {journal} {J. Stat. Mech. Theory Exp.}\ }\textbf {\bibinfo
  {volume} {2009}},\ \bibinfo {pages} {P02025} (\bibinfo {year}
  {2009})}\BibitemShut {NoStop}%
\bibitem [{\citenamefont {Sinitsyn}(2011)}]{Sinitsyn2011}%
  \BibitemOpen
  \bibfield  {author} {\bibinfo {author} {\bibfnamefont {N.~A.}\ \bibnamefont
  {Sinitsyn}},\ }\href@noop {} {\bibfield  {journal} {\bibinfo  {journal} {J.
  Phys. A}\ }\textbf {\bibinfo {volume} {44}},\ \bibinfo {pages} {405001}
  (\bibinfo {year} {2011})}\BibitemShut {NoStop}%
\bibitem [{\citenamefont {Campisi}(2014)}]{Campisi2450}%
  \BibitemOpen
  \bibfield  {author} {\bibinfo {author} {\bibfnamefont {M.}~\bibnamefont
  {Campisi}},\ }\href {\doibase 10.1088/1751-8113/47/24/245001} {\bibfield
  {journal} {\bibinfo  {journal} {J. Phys. A.}\ }\textbf {\bibinfo {volume}
  {47}},\ \bibinfo {pages} {245001} (\bibinfo {year} {2014})}\BibitemShut
  {NoStop}%
\bibitem [{\citenamefont {Garc\'ia-Garc\'ia}\ \emph {et~al.}()\citenamefont
  {Garc\'ia-Garc\'ia}, \citenamefont {Domínguez}, \citenamefont {Lecomte},\
  and\ \citenamefont {Kolton}}]{Garc}%
  \BibitemOpen
  \bibfield  {author} {\bibinfo {author} {\bibfnamefont {R.}~\bibnamefont
  {Garc\'ia-Garc\'ia}}, \bibinfo {author} {\bibfnamefont {D.}~\bibnamefont
  {Domínguez}}, \bibinfo {author} {\bibfnamefont {V.}~\bibnamefont {Lecomte}},
  \ and\ \bibinfo {author} {\bibfnamefont {A.~B.}\ \bibnamefont {Kolton}},\
  }\href {https://doi.org/10.1103/PhysRevE.82.030104} {\bibfield  {journal}
  {\bibinfo  {journal} {Phys. Rev. E}\ }\textbf {\bibinfo {volume} {82}},\
  \bibinfo {pages} {030104(R)}}\BibitemShut {NoStop}%
\bibitem [{\citenamefont {Timpanaro}\ \emph {et~al.}(2019)\citenamefont
  {Timpanaro}, \citenamefont {Guarnieri}, \citenamefont {Goold},\ and\
  \citenamefont {Landi}}]{Guarnieri2019}%
  \BibitemOpen
  \bibfield  {author} {\bibinfo {author} {\bibfnamefont {A.}~\bibnamefont
  {Timpanaro}}, \bibinfo {author} {\bibfnamefont {G.}~\bibnamefont
  {Guarnieri}}, \bibinfo {author} {\bibfnamefont {J.}~\bibnamefont {Goold}}, \
  and\ \bibinfo {author} {\bibfnamefont {G.~T.}\ \bibnamefont {Landi}},\ }\href
  {\doibase 10.1103/PhysRevLett.123.090604} {\bibfield  {journal} {\bibinfo
  {journal} {Phys. Rev. Lett.}\ }\textbf {\bibinfo {volume} {123}},\ \bibinfo
  {pages} {090604} (\bibinfo {year} {2019})}\BibitemShut {NoStop}%
\bibitem [{\citenamefont {Hasegawa}\ and\ \citenamefont
  {Van~Vu}(2019)}]{VuHasegawa2019}%
  \BibitemOpen
  \bibfield  {author} {\bibinfo {author} {\bibfnamefont {Y.}~\bibnamefont
  {Hasegawa}}\ and\ \bibinfo {author} {\bibfnamefont {T.}~\bibnamefont
  {Van~Vu}},\ }\href {\doibase 10.1103/PhysRevLett.123.110602} {\bibfield
  {journal} {\bibinfo  {journal} {Phys. Rev. Lett.}\ }\textbf {\bibinfo
  {volume} {123}},\ \bibinfo {pages} {110602} (\bibinfo {year}
  {2019})}\BibitemShut {NoStop}%
\bibitem [{\citenamefont {Campisi}\ and\ \citenamefont
  {Buffoni}(2020)}]{Campisi2020}%
  \BibitemOpen
  \bibfield  {author} {\bibinfo {author} {\bibfnamefont {M.}~\bibnamefont
  {Campisi}}\ and\ \bibinfo {author} {\bibfnamefont {L.}~\bibnamefont
  {Buffoni}},\ }\href@noop {} {\  (\bibinfo {year} {2020})},\ \Eprint
  {http://arxiv.org/abs/arXiv:2011.01076} {arXiv:2011.01076} \BibitemShut
  {NoStop}%
\bibitem [{\citenamefont {Pietzonka}\ and\ \citenamefont
  {Seifert}(2018)}]{Pietzonka2018}%
  \BibitemOpen
  \bibfield  {author} {\bibinfo {author} {\bibfnamefont {P.}~\bibnamefont
  {Pietzonka}}\ and\ \bibinfo {author} {\bibfnamefont {U.}~\bibnamefont
  {Seifert}},\ }\href {\doibase 10.1103/PhysRevLett.120.190602} {\bibfield
  {journal} {\bibinfo  {journal} {Phys. Rev. Lett.}\ }\textbf {\bibinfo
  {volume} {120}},\ \bibinfo {pages} {190602} (\bibinfo {year}
  {2018})}\BibitemShut {NoStop}%
\bibitem [{\citenamefont {Funo}\ and\ \citenamefont {Ueda}(2015)}]{Funo2018a}%
  \BibitemOpen
  \bibfield  {author} {\bibinfo {author} {\bibfnamefont {K.}~\bibnamefont
  {Funo}}\ and\ \bibinfo {author} {\bibfnamefont {M.}~\bibnamefont {Ueda}},\
  }\href@noop {} {\bibfield  {journal} {\bibinfo  {journal} {Phys. Rev. Lett.}\
  }\textbf {\bibinfo {volume} {115}},\ \bibinfo {pages} {260601} (\bibinfo
  {year} {2015})}\BibitemShut {NoStop}%
\bibitem [{\citenamefont {Holubec}\ and\ \citenamefont
  {Ryabov}(2018)}]{Holubec2018}%
  \BibitemOpen
  \bibfield  {author} {\bibinfo {author} {\bibfnamefont {V.}~\bibnamefont
  {Holubec}}\ and\ \bibinfo {author} {\bibfnamefont {A.}~\bibnamefont
  {Ryabov}},\ }\href {\doibase 10.1103/PhysRevLett.121.120601} {\bibfield
  {journal} {\bibinfo  {journal} {Phys. Rev. Lett.}\ }\textbf {\bibinfo
  {volume} {121}},\ \bibinfo {pages} {120601} (\bibinfo {year}
  {2018})}\BibitemShut {NoStop}%
\bibitem [{\citenamefont {Miller}\ and\ \citenamefont
  {Mehboudi}(2020)}]{Miller2020b}%
  \BibitemOpen
  \bibfield  {author} {\bibinfo {author} {\bibfnamefont {H.}~\bibnamefont
  {Miller}}\ and\ \bibinfo {author} {\bibfnamefont {M.}~\bibnamefont
  {Mehboudi}},\ }\href@noop {} {\  (\bibinfo {year} {2020})},\ \Eprint
  {http://arxiv.org/abs/arXiv:2009.02261} {arXiv:2009.02261} \BibitemShut
  {NoStop}%
\bibitem [{\citenamefont {Campisi}\ \emph {et~al.}(2012)\citenamefont
  {Campisi}, \citenamefont {Denisov},\ and\ \citenamefont
  {H\"anggi}}]{Campisi2012geometric}%
  \BibitemOpen
  \bibfield  {author} {\bibinfo {author} {\bibfnamefont {M.}~\bibnamefont
  {Campisi}}, \bibinfo {author} {\bibfnamefont {S.}~\bibnamefont {Denisov}}, \
  and\ \bibinfo {author} {\bibfnamefont {P.}~\bibnamefont {H\"anggi}},\ }\href
  {\doibase 10.1103/PhysRevA.86.032114} {\bibfield  {journal} {\bibinfo
  {journal} {Phys. Rev. A}\ }\textbf {\bibinfo {volume} {86}},\ \bibinfo
  {pages} {032114} (\bibinfo {year} {2012})}\BibitemShut {NoStop}%
\bibitem [{\citenamefont {Bonan{\c{c}}a}\ and\ \citenamefont
  {Deffner}(2014)}]{bonancca2014optimal}%
  \BibitemOpen
  \bibfield  {author} {\bibinfo {author} {\bibfnamefont {M.~V.~S.}\
  \bibnamefont {Bonan{\c{c}}a}}\ and\ \bibinfo {author} {\bibfnamefont
  {S.}~\bibnamefont {Deffner}},\ }\href
  {https://aip.scitation.org/doi/abs/10.1063/1.4885277?journalCode=jcp}
  {\bibfield  {journal} {\bibinfo  {journal} {The Journal of chemical physics}\
  }\textbf {\bibinfo {volume} {140}},\ \bibinfo {pages} {244119} (\bibinfo
  {year} {2014})}\BibitemShut {NoStop}%
\bibitem [{\citenamefont {Ludovico}\ \emph {et~al.}(2016)\citenamefont
  {Ludovico}, \citenamefont {Battista}, \citenamefont {von Oppen},\ and\
  \citenamefont {Arrachea}}]{Ludovico2016adiabatic}%
  \BibitemOpen
  \bibfield  {author} {\bibinfo {author} {\bibfnamefont {M.~F.}\ \bibnamefont
  {Ludovico}}, \bibinfo {author} {\bibfnamefont {F.}~\bibnamefont {Battista}},
  \bibinfo {author} {\bibfnamefont {F.}~\bibnamefont {von Oppen}}, \ and\
  \bibinfo {author} {\bibfnamefont {L.}~\bibnamefont {Arrachea}},\ }\href
  {\doibase 10.1103/PhysRevB.93.075136} {\bibfield  {journal} {\bibinfo
  {journal} {Phys. Rev. B}\ }\textbf {\bibinfo {volume} {93}},\ \bibinfo
  {pages} {075136} (\bibinfo {year} {2016})}\BibitemShut {NoStop}%
\bibitem [{\citenamefont {Scandi}\ \emph {et~al.}(2020)\citenamefont {Scandi},
  \citenamefont {Miller}, \citenamefont {Anders},\ and\ \citenamefont
  {Perarnau-Llobet}}]{Scandi2020}%
  \BibitemOpen
  \bibfield  {author} {\bibinfo {author} {\bibfnamefont {M.}~\bibnamefont
  {Scandi}}, \bibinfo {author} {\bibfnamefont {H.~J.~D.}\ \bibnamefont
  {Miller}}, \bibinfo {author} {\bibfnamefont {J.}~\bibnamefont {Anders}}, \
  and\ \bibinfo {author} {\bibfnamefont {M.}~\bibnamefont {Perarnau-Llobet}},\
  }\href {\doibase 10.1103/PhysRevResearch.2.023377} {\bibfield  {journal}
  {\bibinfo  {journal} {Phys. Rev. Research}\ }\textbf {\bibinfo {volume}
  {2}},\ \bibinfo {pages} {023377} (\bibinfo {year} {2020})}\BibitemShut
  {NoStop}%
\bibitem [{\citenamefont {Miller}\ \emph {et~al.}(2020)\citenamefont {Miller},
  \citenamefont {Guarnieri}, \citenamefont {Mitchison},\ and\ \citenamefont
  {Goold}}]{Miller2020}%
  \BibitemOpen
  \bibfield  {author} {\bibinfo {author} {\bibfnamefont {H.~J.~D.}\
  \bibnamefont {Miller}}, \bibinfo {author} {\bibfnamefont {G.}~\bibnamefont
  {Guarnieri}}, \bibinfo {author} {\bibfnamefont {M.~T.}\ \bibnamefont
  {Mitchison}}, \ and\ \bibinfo {author} {\bibfnamefont {J.}~\bibnamefont
  {Goold}},\ }\href {\doibase 10.1103/PhysRevLett.125.160602} {\bibfield
  {journal} {\bibinfo  {journal} {Phys. Rev. Lett.}\ }\textbf {\bibinfo
  {volume} {125}},\ \bibinfo {pages} {160602} (\bibinfo {year}
  {2020})}\BibitemShut {NoStop}%
\bibitem [{\citenamefont {Miller}\ \emph {et~al.}(2019)\citenamefont {Miller},
  \citenamefont {Scandi}, \citenamefont {Anders},\ and\ \citenamefont
  {Perarnau-Llobet}}]{Miller2019}%
  \BibitemOpen
  \bibfield  {author} {\bibinfo {author} {\bibfnamefont {H.~J.~D.}\
  \bibnamefont {Miller}}, \bibinfo {author} {\bibfnamefont {M.}~\bibnamefont
  {Scandi}}, \bibinfo {author} {\bibfnamefont {J.}~\bibnamefont {Anders}}, \
  and\ \bibinfo {author} {\bibfnamefont {M.}~\bibnamefont {Perarnau-Llobet}},\
  }\href {\doibase 10.1103/PhysRevLett.123.230603} {\bibfield  {journal}
  {\bibinfo  {journal} {Phys. Rev. Lett.}\ }\textbf {\bibinfo {volume} {123}},\
  \bibinfo {pages} {230603} (\bibinfo {year} {2019})}\BibitemShut {NoStop}%
\bibitem [{\citenamefont {Andrieux}\ \emph {et~al.}(2009)\citenamefont
  {Andrieux}, \citenamefont {Gaspard}, \citenamefont {Monnai},\ and\
  \citenamefont {Tasaki}}]{Andrieux}%
  \BibitemOpen
  \bibfield  {author} {\bibinfo {author} {\bibfnamefont {D.}~\bibnamefont
  {Andrieux}}, \bibinfo {author} {\bibfnamefont {P.}~\bibnamefont {Gaspard}},
  \bibinfo {author} {\bibfnamefont {T.}~\bibnamefont {Monnai}}, \ and\ \bibinfo
  {author} {\bibfnamefont {S.}~\bibnamefont {Tasaki}},\ }\href {\doibase
  10.1088/1367-2630/11/4/043014} {\bibfield  {journal} {\bibinfo  {journal} {N.
  J. Phys}\ }\textbf {\bibinfo {volume} {11}},\ \bibinfo {pages} {043014}
  (\bibinfo {year} {2009})}\BibitemShut {NoStop}%
\bibitem [{\citenamefont {Miller}\ \emph {et~al.}(2021)\citenamefont {Miller},
  \citenamefont {Mohammady}, \citenamefont {Perarnau-Llobet},\ and\
  \citenamefont {Guarnieri}}]{Miller2020a}%
  \BibitemOpen
  \bibfield  {author} {\bibinfo {author} {\bibfnamefont {H.~J.~D.}\
  \bibnamefont {Miller}}, \bibinfo {author} {\bibfnamefont {M.~H.}\
  \bibnamefont {Mohammady}}, \bibinfo {author} {\bibfnamefont {M.}~\bibnamefont
  {Perarnau-Llobet}}, \ and\ \bibinfo {author} {\bibfnamefont {G.}~\bibnamefont
  {Guarnieri}},\ }\href {\doibase 10.1103/PhysRevLett.126.210603} {\bibfield
  {journal} {\bibinfo  {journal} {Phys. Rev. Lett.}\ }\textbf {\bibinfo
  {volume} {126}},\ \bibinfo {pages} {210603} (\bibinfo {year}
  {2021})}\BibitemShut {NoStop}%
\bibitem [{\citenamefont {Mandal}\ and\ \citenamefont
  {Jarzynski}(2016)}]{Mandal2016a}%
  \BibitemOpen
  \bibfield  {author} {\bibinfo {author} {\bibfnamefont {D.}~\bibnamefont
  {Mandal}}\ and\ \bibinfo {author} {\bibfnamefont {C.}~\bibnamefont
  {Jarzynski}},\ }\href {\doibase 10.1088/1742-5468/2016/06/063204} {\bibfield
  {journal} {\bibinfo  {journal} {J. Stat. Mech.}\ }\textbf {\bibinfo {volume}
  {2016}},\ \bibinfo {pages} {063204} (\bibinfo {year} {2016})}\BibitemShut
  {NoStop}%
\bibitem [{\citenamefont {Esposito}\ and\ \citenamefont {{Van Den
  Broeck}}(2010)}]{Esposito2010b}%
  \BibitemOpen
  \bibfield  {author} {\bibinfo {author} {\bibfnamefont {M.}~\bibnamefont
  {Esposito}}\ and\ \bibinfo {author} {\bibfnamefont {C.}~\bibnamefont {{Van
  Den Broeck}}},\ }\href {\doibase 10.1103/PhysRevLett.104.090601} {\bibfield
  {journal} {\bibinfo  {journal} {Phys. Rev. Lett.}\ }\textbf {\bibinfo
  {volume} {104}},\ \bibinfo {pages} {090601} (\bibinfo {year}
  {2010})}\BibitemShut {NoStop}%
\bibitem [{\citenamefont {Esposito}\ and\ \citenamefont {{Van den
  Broeck}}(2010)}]{Esposito2018}%
  \BibitemOpen
  \bibfield  {author} {\bibinfo {author} {\bibfnamefont {M.}~\bibnamefont
  {Esposito}}\ and\ \bibinfo {author} {\bibfnamefont {C.}~\bibnamefont {{Van
  den Broeck}}},\ }\href {\doibase 10.1103/PhysRevE.82.011143} {\bibfield
  {journal} {\bibinfo  {journal} {Phys. Rev. E}\ }\textbf {\bibinfo {volume}
  {82}},\ \bibinfo {pages} {011143} (\bibinfo {year} {2010})}\BibitemShut
  {NoStop}%
\bibitem [{\citenamefont {Albash}\ \emph {et~al.}(2012)\citenamefont {Albash},
  \citenamefont {Boixo},\ and\ \citenamefont {Lidar}}]{Albash2012}%
  \BibitemOpen
  \bibfield  {author} {\bibinfo {author} {\bibfnamefont {T.}~\bibnamefont
  {Albash}}, \bibinfo {author} {\bibfnamefont {S.}~\bibnamefont {Boixo}}, \
  and\ \bibinfo {author} {\bibfnamefont {D.~A.}\ \bibnamefont {Lidar}},\ }\href
  {\doibase 10.1088/1367-2630/14/12/123016} {\bibfield  {journal} {\bibinfo
  {journal} {N. J. Phys}\ }\textbf {\bibinfo {volume} {14}},\ \bibinfo {pages}
  {123016} (\bibinfo {year} {2012})}\BibitemShut {NoStop}%
\bibitem [{\citenamefont {Rivas}\ and\ \citenamefont
  {Huelga}(2012)}]{Rivas2011}%
  \BibitemOpen
  \bibfield  {author} {\bibinfo {author} {\bibfnamefont {A.}~\bibnamefont
  {Rivas}}\ and\ \bibinfo {author} {\bibfnamefont {S.}~\bibnamefont {Huelga}},\
  }\href {\doibase 10.1007/978-3-642-23354-8} {\emph {\bibinfo {title} {{Open
  Quantum Systems: An Introduction}}}}\ (\bibinfo  {publisher} {Springer
  Heidelberg},\ \bibinfo {year} {2012})\BibitemShut {NoStop}%
\bibitem [{\citenamefont {Hatano}\ and\ \citenamefont
  {Sasa}(2001)}]{Hatano2001b}%
  \BibitemOpen
  \bibfield  {author} {\bibinfo {author} {\bibfnamefont {T.}~\bibnamefont
  {Hatano}}\ and\ \bibinfo {author} {\bibfnamefont {S.}~\bibnamefont {Sasa}},\
  }\href {\doibase 10.1103/PhysRevLett.86.3463} {\bibfield  {journal} {\bibinfo
   {journal} {Phys. Rev. Lett.}\ }\textbf {\bibinfo {volume} {86}},\ \bibinfo
  {pages} {3463} (\bibinfo {year} {2001})}\BibitemShut {NoStop}%
\bibitem [{\citenamefont {Alicki}\ and\ \citenamefont
  {Lendi}(2007)}]{Holevo2001}%
  \BibitemOpen
  \bibfield  {author} {\bibinfo {author} {\bibfnamefont {R.}~\bibnamefont
  {Alicki}}\ and\ \bibinfo {author} {\bibfnamefont {K.}~\bibnamefont {Lendi}},\
  }\href@noop {} {\emph {\bibinfo {title} {{Quantum Dynamical Semigroups and
  Applications}}}}\ (\bibinfo  {publisher} {Springer},\ \bibinfo {year}
  {2007})\BibitemShut {NoStop}%
\bibitem [{\citenamefont {Norris}(1997)}]{Norris1997}%
  \BibitemOpen
  \bibfield  {author} {\bibinfo {author} {\bibfnamefont {J.~R.}\ \bibnamefont
  {Norris}},\ }\href@noop {} {\emph {\bibinfo {title} {{Markov Chains}}}}\
  (\bibinfo  {publisher} {Cambridge University Press},\ \bibinfo {address}
  {Cambridge},\ \bibinfo {year} {1997})\BibitemShut {NoStop}%
\bibitem [{\citenamefont {Crooks}(2008)}]{Crooks2018}%
  \BibitemOpen
  \bibfield  {author} {\bibinfo {author} {\bibfnamefont {G.~E.}\ \bibnamefont
  {Crooks}},\ }\href {\doibase 10.1103/PhysRevA.77.034101} {\bibfield
  {journal} {\bibinfo  {journal} {Phys. Rev. A}\ }\textbf {\bibinfo {volume}
  {77}},\ \bibinfo {pages} {034101(4)} (\bibinfo {year} {2008})}\BibitemShut
  {NoStop}%
\bibitem [{\citenamefont {Fagnola}\ and\ \citenamefont
  {Umanita}(2007)}]{Fagnola2007}%
  \BibitemOpen
  \bibfield  {author} {\bibinfo {author} {\bibfnamefont {F.}~\bibnamefont
  {Fagnola}}\ and\ \bibinfo {author} {\bibfnamefont {V.}~\bibnamefont
  {Umanita}},\ }\href {\doibase 10.1142/S0219025707002762} {\bibfield
  {journal} {\bibinfo  {journal} {Infin. Dimens. Anal. Quantum Probab. Relat.
  Top.}\ }\textbf {\bibinfo {volume} {10}},\ \bibinfo {pages} {335} (\bibinfo
  {year} {2007})}\BibitemShut {NoStop}%
\bibitem [{\citenamefont {Alicki}(1976)}]{Alicki1976}%
  \BibitemOpen
  \bibfield  {author} {\bibinfo {author} {\bibfnamefont {R.}~\bibnamefont
  {Alicki}},\ }\href {\doibase 10.1016/0034-4877(76)90046-X} {\bibfield
  {journal} {\bibinfo  {journal} {Rep. Math. Phys.}\ }\textbf {\bibinfo
  {volume} {10}},\ \bibinfo {pages} {249} (\bibinfo {year} {1976})}\BibitemShut
  {NoStop}%
\bibitem [{\citenamefont {Brandner}\ and\ \citenamefont
  {Saito}(2020)}]{Brandner2020}%
  \BibitemOpen
  \bibfield  {author} {\bibinfo {author} {\bibfnamefont {K.}~\bibnamefont
  {Brandner}}\ and\ \bibinfo {author} {\bibfnamefont {K.}~\bibnamefont
  {Saito}},\ }\href {\doibase 10.1103/PhysRevLett.124.040602} {\bibfield
  {journal} {\bibinfo  {journal} {Phys. Rev. Lett.}\ }\textbf {\bibinfo
  {volume} {124}},\ \bibinfo {pages} {040602} (\bibinfo {year}
  {2020})}\BibitemShut {NoStop}%
\bibitem [{\citenamefont {Sivak}\ and\ \citenamefont
  {Crooks}(2012)}]{Sivak2012a}%
  \BibitemOpen
  \bibfield  {author} {\bibinfo {author} {\bibfnamefont {D.~A.}\ \bibnamefont
  {Sivak}}\ and\ \bibinfo {author} {\bibfnamefont {G.~E.}\ \bibnamefont
  {Crooks}},\ }\href {\doibase 10.1103/PhysRevLett.108.190602} {\bibfield
  {journal} {\bibinfo  {journal} {Phys. Rev. L}\ }\textbf {\bibinfo {volume}
  {108}},\ \bibinfo {pages} {190602 (2012)} (\bibinfo {year}
  {2012})}\BibitemShut {NoStop}%
\bibitem [{\citenamefont {Scandi}\ and\ \citenamefont
  {Perarnau-Llobet}(2019)}]{Scandi}%
  \BibitemOpen
  \bibfield  {author} {\bibinfo {author} {\bibfnamefont {M.}~\bibnamefont
  {Scandi}}\ and\ \bibinfo {author} {\bibfnamefont {M.}~\bibnamefont
  {Perarnau-Llobet}},\ }\href {\doibase doi.org/10.22331/q-2019-10-24-197}
  {\bibfield  {journal} {\bibinfo  {journal} {Quantum}\ }\textbf {\bibinfo
  {volume} {3}},\ \bibinfo {pages} {197} (\bibinfo {year} {2019})}\BibitemShut
  {NoStop}%
\bibitem [{\citenamefont {Znidaric}(2015)}]{Znidari2015}%
  \BibitemOpen
  \bibfield  {author} {\bibinfo {author} {\bibfnamefont {M.}~\bibnamefont
  {Znidaric}},\ }\href@noop {} {\bibfield  {journal} {\bibinfo  {journal}
  {Phys. Rev. E}\ }\textbf {\bibinfo {volume} {92}},\ \bibinfo {pages} {042143}
  (\bibinfo {year} {2015})}\BibitemShut {NoStop}%
\bibitem [{\citenamefont {Boullion}\ and\ \citenamefont
  {Odell}(1971)}]{Boullion1971}%
  \BibitemOpen
  \bibfield  {author} {\bibinfo {author} {\bibfnamefont {T.~L.}\ \bibnamefont
  {Boullion}}\ and\ \bibinfo {author} {\bibfnamefont {P.~L.}\ \bibnamefont
  {Odell}},\ }\href@noop {} {\emph {\bibinfo {title} {{Generalised inverse
  matrices}}}}\ (\bibinfo  {publisher} {Wiley-Interscience},\ \bibinfo
  {address} {New York},\ \bibinfo {year} {1971})\BibitemShut {NoStop}%
\bibitem [{\citenamefont {Cavina}\ \emph {et~al.}(2017)\citenamefont {Cavina},
  \citenamefont {Mari},\ and\ \citenamefont {Giovannetti}}]{Cavina2017}%
  \BibitemOpen
  \bibfield  {author} {\bibinfo {author} {\bibfnamefont {V.}~\bibnamefont
  {Cavina}}, \bibinfo {author} {\bibfnamefont {A.}~\bibnamefont {Mari}}, \ and\
  \bibinfo {author} {\bibfnamefont {V.}~\bibnamefont {Giovannetti}},\ }\href
  {\doibase 10.1103/PhysRevLett.119.050601} {\bibfield  {journal} {\bibinfo
  {journal} {Phys. Rev. Lett.}\ }\textbf {\bibinfo {volume} {119}},\ \bibinfo
  {pages} {050601} (\bibinfo {year} {2017})}\BibitemShut {NoStop}%
\bibitem [{\citenamefont {Petz}(2002)}]{Petz2002}%
  \BibitemOpen
  \bibfield  {author} {\bibinfo {author} {\bibfnamefont {D.}~\bibnamefont
  {Petz}},\ }\href {\doibase 10.1088/0305-4470/35/4/305} {\bibfield  {journal}
  {\bibinfo  {journal} {J. Phys. A}\ }\textbf {\bibinfo {volume} {35}},\
  \bibinfo {pages} {929} (\bibinfo {year} {2002})}\BibitemShut {NoStop}%
\bibitem [{\citenamefont {Evans}\ and\ \citenamefont {Searles}(2002)}]{E2002}%
  \BibitemOpen
  \bibfield  {author} {\bibinfo {author} {\bibfnamefont {D.~J.}\ \bibnamefont
  {Evans}}\ and\ \bibinfo {author} {\bibfnamefont {D.~J.}\ \bibnamefont
  {Searles}},\ }\href {\doibase 10.1080/00018730210155133} {\bibfield
  {journal} {\bibinfo  {journal} {Adv. Phys.}\ }\textbf {\bibinfo {volume}
  {51}},\ \bibinfo {pages} {1529} (\bibinfo {year} {2002})}\BibitemShut
  {NoStop}%
\bibitem [{\citenamefont {Jarzynski}\ and\ \citenamefont
  {Wojcik}(2004)}]{Jarzynski2004b}%
  \BibitemOpen
  \bibfield  {author} {\bibinfo {author} {\bibfnamefont {C.}~\bibnamefont
  {Jarzynski}}\ and\ \bibinfo {author} {\bibfnamefont {D.~K.}\ \bibnamefont
  {Wojcik}},\ }\href {\doibase 10.1103/PhysRevLett.92.230602} {\bibfield
  {journal} {\bibinfo  {journal} {Phys. Rev. Lett.}\ }\textbf {\bibinfo
  {volume} {92}},\ \bibinfo {pages} {230602} (\bibinfo {year}
  {2004})}\BibitemShut {NoStop}%
\bibitem [{\citenamefont {Jeon}\ \emph {et~al.}(2017)\citenamefont {Jeon},
  \citenamefont {Talkner}, \citenamefont {Yi},\ and\ \citenamefont
  {Woon~Kim}}]{Jeon2017}%
  \BibitemOpen
  \bibfield  {author} {\bibinfo {author} {\bibfnamefont {E.}~\bibnamefont
  {Jeon}}, \bibinfo {author} {\bibfnamefont {P.}~\bibnamefont {Talkner}},
  \bibinfo {author} {\bibfnamefont {J.}~\bibnamefont {Yi}}, \ and\ \bibinfo
  {author} {\bibfnamefont {Y.}~\bibnamefont {Woon~Kim}},\ }\href {\doibase
  10.1088/1367-2630/aa8110} {\bibfield  {journal} {\bibinfo  {journal} {N. J.
  Phys}\ }\textbf {\bibinfo {volume} {19}},\ \bibinfo {pages} {093006}
  (\bibinfo {year} {2017})}\BibitemShut {NoStop}%
\bibitem [{\citenamefont {Crooks}(1999)}]{Crooks}%
  \BibitemOpen
  \bibfield  {author} {\bibinfo {author} {\bibfnamefont {G.}~\bibnamefont
  {Crooks}},\ }\href {\doibase 10.1103/PhysRevE.60.2721} {\bibfield  {journal}
  {\bibinfo  {journal} {Phys. Rev. E}\ }\textbf {\bibinfo {volume} {60}},\
  \bibinfo {pages} {2721} (\bibinfo {year} {1999})}\BibitemShut {NoStop}%
\bibitem [{\citenamefont {Merhav}\ and\ \citenamefont
  {Kafri}(2010)}]{Merhav2010}%
  \BibitemOpen
  \bibfield  {author} {\bibinfo {author} {\bibfnamefont {N.}~\bibnamefont
  {Merhav}}\ and\ \bibinfo {author} {\bibfnamefont {Y.}~\bibnamefont {Kafri}},\
  }\href {\doibase 10.1088/1742-5468/2010/12/P12022} {\bibfield  {journal}
  {\bibinfo  {journal} {J. Stat. Mech.}\ }\textbf {\bibinfo {volume} {2010}},\
  \bibinfo {pages} {P12022} (\bibinfo {year} {2010})}\BibitemShut {NoStop}%
\bibitem [{\citenamefont {Perarnau-Llobet}\ \emph {et~al.}(2016)\citenamefont
  {Perarnau-Llobet}, \citenamefont {Riera}, \citenamefont {Gallego},
  \citenamefont {Wilming},\ and\ \citenamefont
  {Eisert}}]{Perarnau-Llobet2016a}%
  \BibitemOpen
  \bibfield  {author} {\bibinfo {author} {\bibfnamefont {M.}~\bibnamefont
  {Perarnau-Llobet}}, \bibinfo {author} {\bibfnamefont {A.}~\bibnamefont
  {Riera}}, \bibinfo {author} {\bibfnamefont {R.}~\bibnamefont {Gallego}},
  \bibinfo {author} {\bibfnamefont {H.}~\bibnamefont {Wilming}}, \ and\
  \bibinfo {author} {\bibfnamefont {J.}~\bibnamefont {Eisert}},\ }\href
  {\doibase 10.1088/1367-2630/aa4fa6} {\bibfield  {journal} {\bibinfo
  {journal} {N. J. Phys}\ }\textbf {\bibinfo {volume} {18}},\ \bibinfo {pages}
  {1} (\bibinfo {year} {2016})}\BibitemShut {NoStop}%
\bibitem [{\citenamefont {Manzano}\ \emph {et~al.}(2016)\citenamefont
  {Manzano}, \citenamefont {Galve}, \citenamefont {Zambrini},\ and\
  \citenamefont {Parrondo}}]{Manzano2016}%
  \BibitemOpen
  \bibfield  {author} {\bibinfo {author} {\bibfnamefont {G.}~\bibnamefont
  {Manzano}}, \bibinfo {author} {\bibfnamefont {F.}~\bibnamefont {Galve}},
  \bibinfo {author} {\bibfnamefont {R.}~\bibnamefont {Zambrini}}, \ and\
  \bibinfo {author} {\bibfnamefont {J.~M.}\ \bibnamefont {Parrondo}},\ }\href
  {\doibase 10.1103/PhysRevE.93.052120} {\bibfield  {journal} {\bibinfo
  {journal} {Phys. Rev. E}\ }\textbf {\bibinfo {volume} {93}},\ \bibinfo
  {pages} {052120} (\bibinfo {year} {2016})}\BibitemShut {NoStop}%
\bibitem [{\citenamefont {Petz}\ and\ \citenamefont
  {Szab{\'o}}(2009)}]{DenesPetz2009}%
  \BibitemOpen
  \bibfield  {author} {\bibinfo {author} {\bibfnamefont {D.}~\bibnamefont
  {Petz}}\ and\ \bibinfo {author} {\bibfnamefont {V.~S.}\ \bibnamefont
  {Szab{\'o}}},\ }\href@noop {} {\bibfield  {journal} {\bibinfo  {journal}
  {International Journal of Mathematics}\ }\textbf {\bibinfo {volume} {20}},\
  \bibinfo {pages} {1421} (\bibinfo {year} {2009})}\BibitemShut {NoStop}%
\bibitem [{\citenamefont {Hansen}(2008)}]{Hansen2008}%
  \BibitemOpen
  \bibfield  {author} {\bibinfo {author} {\bibfnamefont {F.}~\bibnamefont
  {Hansen}},\ }\href {\doibase 10.1073/pnas.0803323105} {\bibfield  {journal}
  {\bibinfo  {journal} {Proc. Natl. Acad. Sci. USA}\ }\textbf {\bibinfo
  {volume} {105}},\ \bibinfo {pages} {9909} (\bibinfo {year}
  {2008})}\BibitemShut {NoStop}%
\bibitem [{\citenamefont {Feldmann}\ and\ \citenamefont
  {Kosloff}(2003)}]{Feldmann}%
  \BibitemOpen
  \bibfield  {author} {\bibinfo {author} {\bibfnamefont {T.}~\bibnamefont
  {Feldmann}}\ and\ \bibinfo {author} {\bibfnamefont {R.}~\bibnamefont
  {Kosloff}},\ }\href {\doibase 10.1103/PhysRevE.68.016101} {\bibfield
  {journal} {\bibinfo  {journal} {Phys. Rev. E}\ }\textbf {\bibinfo {volume}
  {68}},\ \bibinfo {pages} {016101} (\bibinfo {year} {2003})}\BibitemShut
  {NoStop}%
\bibitem [{\citenamefont {Plastina}\ \emph {et~al.}(2014)\citenamefont
  {Plastina}, \citenamefont {Alecce}, \citenamefont {Apollaro}, \citenamefont
  {Falcone}, \citenamefont {Francica}, \citenamefont {Galve}, \citenamefont
  {Gullo},\ and\ \citenamefont {Zambrini}}]{Plastina}%
  \BibitemOpen
  \bibfield  {author} {\bibinfo {author} {\bibfnamefont {F.}~\bibnamefont
  {Plastina}}, \bibinfo {author} {\bibfnamefont {A.}~\bibnamefont {Alecce}},
  \bibinfo {author} {\bibfnamefont {T.~J.~G.}\ \bibnamefont {Apollaro}},
  \bibinfo {author} {\bibfnamefont {G.}~\bibnamefont {Falcone}}, \bibinfo
  {author} {\bibfnamefont {G.}~\bibnamefont {Francica}}, \bibinfo {author}
  {\bibfnamefont {F.}~\bibnamefont {Galve}}, \bibinfo {author} {\bibfnamefont
  {N.~L.}\ \bibnamefont {Gullo}}, \ and\ \bibinfo {author} {\bibfnamefont
  {R.}~\bibnamefont {Zambrini}},\ }\href {\doibase
  10.1103/PhysRevLett.113.260601} {\bibfield  {journal} {\bibinfo  {journal}
  {Phys. Rev. Lett.}\ }\textbf {\bibinfo {volume} {113}},\ \bibinfo {pages}
  {260601} (\bibinfo {year} {2014})}\BibitemShut {NoStop}%
\bibitem [{\citenamefont {Brandner}\ \emph {et~al.}(2017)\citenamefont
  {Brandner}, \citenamefont {Bauer},\ and\ \citenamefont {Seifert}}]{Brandner}%
  \BibitemOpen
  \bibfield  {author} {\bibinfo {author} {\bibfnamefont {K.}~\bibnamefont
  {Brandner}}, \bibinfo {author} {\bibfnamefont {M.}~\bibnamefont {Bauer}}, \
  and\ \bibinfo {author} {\bibfnamefont {U.}~\bibnamefont {Seifert}},\ }\href
  {\doibase 10.1103/PhysRevLett.119.170602} {\bibfield  {journal} {\bibinfo
  {journal} {Phys. Rev. Lett.}\ }\textbf {\bibinfo {volume} {119}},\ \bibinfo
  {pages} {170602} (\bibinfo {year} {2017})}\BibitemShut {NoStop}%
\bibitem [{\citenamefont {Abiuso}\ and\ \citenamefont
  {Perarnau-Llobet}(2020)}]{Abiuso2020}%
  \BibitemOpen
  \bibfield  {author} {\bibinfo {author} {\bibfnamefont {P.}~\bibnamefont
  {Abiuso}}\ and\ \bibinfo {author} {\bibfnamefont {M.}~\bibnamefont
  {Perarnau-Llobet}},\ }\href {\doibase 10.1103/PhysRevLett.124.110606}
  {\bibfield  {journal} {\bibinfo  {journal} {Phys. Rev. Lett.}\ }\textbf
  {\bibinfo {volume} {124}},\ \bibinfo {pages} {110606} (\bibinfo {year}
  {2020})}\BibitemShut {NoStop}%
\bibitem [{\citenamefont {Abiuso}\ \emph {et~al.}(2020)\citenamefont {Abiuso},
  \citenamefont {Miller}, \citenamefont {Perarnau-Llobet},\ and\ \citenamefont
  {Scandi}}]{Abiuso2020a}%
  \BibitemOpen
  \bibfield  {author} {\bibinfo {author} {\bibfnamefont {P.}~\bibnamefont
  {Abiuso}}, \bibinfo {author} {\bibfnamefont {H.~J.~D.}\ \bibnamefont
  {Miller}}, \bibinfo {author} {\bibfnamefont {M.}~\bibnamefont
  {Perarnau-Llobet}}, \ and\ \bibinfo {author} {\bibfnamefont {M.}~\bibnamefont
  {Scandi}},\ }\href {\doibase 10.3390/e22101076} {\bibfield  {journal}
  {\bibinfo  {journal} {Entropy}\ }\textbf {\bibinfo {volume} {22}},\ \bibinfo
  {pages} {1076} (\bibinfo {year} {2020})}\BibitemShut {NoStop}%
\bibitem [{\citenamefont {Speck}\ and\ \citenamefont {Seifert}(2004)}]{Speck}%
  \BibitemOpen
  \bibfield  {author} {\bibinfo {author} {\bibfnamefont {T.}~\bibnamefont
  {Speck}}\ and\ \bibinfo {author} {\bibfnamefont {U.}~\bibnamefont
  {Seifert}},\ }\href@noop {} {\bibfield  {journal} {\bibinfo  {journal} {Phys.
  Rev. E}\ }\textbf {\bibinfo {volume} {70}},\ \bibinfo {pages} {066112}
  (\bibinfo {year} {2004})}\BibitemShut {NoStop}%
\bibitem [{\citenamefont {Fagnola}\ and\ \citenamefont
  {Quezada}(2005)}]{Fagnola-harmonic-oscillator}%
  \BibitemOpen
  \bibfield  {author} {\bibinfo {author} {\bibfnamefont {F.}~\bibnamefont
  {Fagnola}}\ and\ \bibinfo {author} {\bibfnamefont {R.}~\bibnamefont
  {Quezada}},\ }\href {\doibase 10.1142/S0219025705002116} {\bibfield
  {journal} {\bibinfo  {journal} {Infin. Dimens. Anal. Quantum Probab. Relat.
  Top.}\ }\textbf {\bibinfo {volume} {08}},\ \bibinfo {pages} {573} (\bibinfo
  {year} {2005})}\BibitemShut {NoStop}%
\bibitem [{\citenamefont {Petz}\ and\ \citenamefont {Hiai}(2014)}]{Petz2014}%
  \BibitemOpen
  \bibfield  {author} {\bibinfo {author} {\bibfnamefont {D.}~\bibnamefont
  {Petz}}\ and\ \bibinfo {author} {\bibfnamefont {F.}~\bibnamefont {Hiai}},\
  }\href@noop {} {\emph {\bibinfo {title} {{Introduction to Matrix Analysis and
  Applications}}}}\ (\bibinfo  {publisher} {Springer-Verlag},\ \bibinfo
  {address} {New Delhi},\ \bibinfo {year} {2014})\BibitemShut {NoStop}%
\end{thebibliography}%

\appendix

\widetext

\section{Proof of~\eqref{eq:priv_ident}}\label{app:1}

Let $p_0>0$ denote the smallest eigenvalue of $\pi_\lambda$. For any $u\in\mathbb{R}$ the real-valued function $f(z)=z^u$ is continuous on the bounded interval $[p_0,1]$, and so by the Stone-Weierstrass theorem can be uniformly approximated by a polynomial $f(z)=\sum_k c_k z^k$ on $[p_0,1]$. For a polynomial we find
\begin{align}
    \nonumber f(\pi_\lambda)K_x(\lambda) &=\sum_k c_k \pi_\lambda^k K_x(\lambda), \\
    \nonumber&=K_x(\lambda)\sum_k c_k e^{-k \Delta \phi_x(\lambda)}\pi_\lambda^k, \\
    \nonumber&=K_x(\lambda)\sum_k c_k (e^{-\Delta \phi_x(\lambda)}\pi_\lambda)^k, \\
    &=K_x(\lambda) f(e^{-\Delta \phi_x(\lambda)}\pi_\lambda),
\end{align}
where we used~\eqref{eq:priv2} in the second line. Since $f(z)=z^u$ we therefore have 
\begin{align}
    \pi_\lambda^u K_x(\lambda)=e^{-u\Delta \phi_x(\lambda)}K_x(\lambda)\pi_\lambda^{u}.
\end{align}

\section{Derivation of~\eqref{eq:MGFexact}}\label{app:b2}

Let us first define the evolution map $e^{\delta t \lind_{\lambda_n}}=\sum_{x_n} K_{x_n}(\lambda_n) (\cdot)K_{x_n}^\dagger(\lambda_n)$, with $\delta t:= \max|t_{n+1}-t_n|$, for the components of the product~\eqref{eq:split}, and recall the steady-state boundary conditions \eq{eq:boundary},  $\rho_0 = \pi_{\lambda_0} = \sum_\mu p_\mu(\lambda_0) \Pi_\mu(\lambda_0)$ and $\rho_\tau =  \pi_{\lambda_{N+1}} = \sum_\nu p_\nu(\lambda_{N+1}) \Pi_{\nu}(\lambda_{N+1})$. Therefore, defining  $y_n := u+T_n v$, and  combining~\eqref{eq:trajecprob-app} with~\eqref{eq:entprod_thermal},~\eqref{eq:tilde-diss-app} and~\eqref{eq:mgf1}, we obtain
\begin{align}\label{eq:mgf2}
   \mathcal{G}_{\sigma,\tilde w}(u,v)&=  \sum_{\mu,\nu}\sum_{\{ x_n \}} \tr{\Pi_\nu (\lambda_{N+1})K_{x_N}(\lambda_N)\dots K_{x_0}(\lambda_0) \Pi_\mu(\lambda_0) K_{x_0}^\dagger(\lambda_0) \dots K_{x_N}^\dagger(\lambda_N)} \nonumber \\
   & \qquad \qquad \qquad p_\nu^{y_{N+1}}(\lambda_{N+1}) p_\mu^{1-y_0}(\lambda_0) e^{y_0 \Delta \phi_{x_0}(\lambda_0)}\dots e^{y_N \Delta \phi_{x_N}(\lambda_N)}e^{v (\ww-\Delta F)}, \nonumber \\
   & = e^{v(\ww-\Delta F)}\sum_{\{ x_n \}} \text{Tr}\bigg(\pi_{\lambda_{N+1}}^{y_{N+1}} (\pi_{\lambda_N}^{-{y_N}/2 } K_{x_N}(\lambda_N) \pi_{\lambda_N}^{y_N/2})\dots \nonumber \\
   & \qquad \qquad \qquad \qquad \qquad \qquad \dots (\pi_{\lambda_0}^{-{y_0}/2 } K_{x_0}(\lambda_0) \pi_{\lambda_0}^{y_0/2}) \pi_{\lambda_0}^{1-y_0} (\pi_{\lambda_0}^{y_0/2 } K_{x_0}^\dagger(\lambda_0) \pi_{\lambda_0}^{-y_0/2}) \dots \nonumber \\
   \nonumber&  \ \ \ \ \ \ \ \ \ \ \ \ \ \ \ \ \ \ \ \ \ \ \ \ \ \ \ \ \ \ \ \ \ \ \ \ \ \ \ \ \ \ \ \ \ \ \ \ \ \ \ \ \ \ \ \ \ \ \ \ \ \ \ \ \ \ \ \ \ \ \ \ \ \ \ \ \ \ \ \ \ \ \ \ \ \ \ \  \dots (\pi_{\lambda_N}^{y_N/2 } K_{x_N}^\dagger(\lambda_N) \pi_{\lambda_N}^{-y_N/2})\bigg),  \\
   & = e^{v(\ww-\Delta F)}\sum_{\{ x_n \}} \text{Tr}\bigg((\pi_{\lambda(N+1)}^{y_{N+1}/2} \pi_{\lambda_N}^{-{y_N}/2 } K_{x_N}(\lambda_N))\dots \nonumber \\
   & \qquad \qquad \qquad \qquad \qquad \qquad \dots (\pi_{\lambda_0}^{{y_1}/2 }\pi_{\lambda_0}^{-{y_0}/2 } K_{x_0}(\lambda_0) ) \pi_{\lambda_0} (K_{x_0}^\dagger(\lambda_0) \pi_{\lambda_0}^{-y_0/2} \pi_{\lambda_0}^{{y_1}/2 }) \dots \nonumber \\
   \nonumber&  \ \ \ \ \ \ \ \ \ \ \ \ \ \ \ \ \ \ \ \ \ \ \ \ \ \ \ \ \ \ \ \ \ \ \ \ \ \ \ \ \ \ \ \ \ \ \ \ \ \ \ \ \ \ \ \ \ \ \ \ \ \ \ \ \ \ \ \ \ \ \ \ \ \ \ \ \ \ \ \ \ \ \ \ \ \ \ \  \dots ( K_{x_N}^\dagger (\lambda_N) \pi_{\lambda_N}^{-y_N/2} \pi_{\lambda_{N+1}}^{y_{N+1}/2}) \bigg),  \\
   &= e^{v(\ww-\Delta F)} \tr{\bigg(\prod^0_{n=N}\mathcal{M}_{u,v}^{(n)}\bigg)(\pi_{\lambda_0})},
\end{align}
where we used $\pi^u_{\lambda(t)} =\sum_\mu p^u_\mu(\lambda(t)) \Pi_\mu(\lambda(t))$,  made use of the privileged representation from~\eqref{eq:priv_ident},  and introduced the linear map
\begin{align}\label{eq:map1}
    \mathcal{M}_{u,v}^{(n)}(\cdot):=\pi^{(u+T_{n+1}v)/2}_{\lambda_{n+1}}\pi^{-(u+T_{n}v)/2}_{\lambda_{n}}e^{\delta t \lind_{\lambda_n}}(\cdot)\pi^{-(u+T_{n}v)/2}_{\lambda_{n}}\pi^{(u+T_{n+1}v)/2}_{\lambda_{n+1}}.
\end{align}
We next utilise the Taylor expansion of the exponential operator for $X,Y\in\bh$ \cite{Petz2014}:
\begin{align}\label{eq:Taylor-Exp}
    e^{-X-\delta t Y} e^X=\id-\delta t\int^1_0 ds \ e^{-sX}Ye^{sX}+\mathcal{O}(\delta t^2).
\end{align}
 For convenience let us introduce the notation $ \mathscr{H}_{n}^{(u,v)}\delta t= (u\beta_{n+1} + v)\tilde{H}_{\lambda_{n+1}}-(u\beta_n + v)\tilde{H}_{\lambda_{n}}$ with $\tilde{H}_\lambda=H_{\Lambda}-F_\lambda \id$. Since the steady state is thermal, i.e. $\pi_\lambda=e^{-\beta \tilde H_\lambda}$, by using~\eqref{eq:Taylor-Exp} we obtain
 \begin{align}
    \nonumber\pi^{(u+T_{n+1}v)/2}_{\lambda_{n+1}}\pi^{-(u+T_{n}v)/2}_{\lambda_{n}}&=\id-\frac{ \delta t}{2} \int^1_0 ds \ e^{-s(u\beta_n+v)\frac{\tilde{H}_{\lambda_{n}}}{2}} \mathscr{H}^{(u,v)}_n e^{s(u\beta_n+v)\frac{\tilde{H}_{\lambda_{n}}}{2}}+\mathcal{O}(\delta t^2), \\
    &=\exp\bigg(-\frac{ \delta t}{2} \int^1_0 ds \ e^{-s(u\beta_n+v)\frac{\tilde{H}_{\lambda_{n}}}{2}} \mathscr{H}^{(u,v)}_n e^{s(u\beta_n+v)\frac{\tilde{H}_{\lambda_{n}}}{2}} +\mathcal{O}(\delta t^2)\bigg).
 \end{align}
Using the fact that $\pi^{-(u+T_{n}v)/2}_{\lambda_{n}}\pi^{(u+T_{n+1}v)/2}_{\lambda_{n+1}}=\big(\pi^{(u+T_{n+1}v)/2}_{\lambda_{n+1}}\pi^{-(u+T_{n}v)/2}_{\lambda_{n}}\big)^\dagger$ with $e^{X}(\cdot)e^{X^\dagger} := \exp(X\star)$, and that the terms in~\eqref{eq:map1} commute to first order in $\delta t$, we have
\begin{align}\label{eq:limit1}
    \mathcal{M}_{u,v}^{(n)}(\cdot)=\exp\bigg(\delta t \lind_{\lambda_n}-\frac{ \delta t}{2} \int^1_0 ds \ e^{-s(u\beta_n+v)\tilde{H}_{\lambda_n}/2}  \mathscr{H}^{(u,v)}_n e^{s(u\beta_n+v)\tilde{H}_{\lambda_n}/2}\star +\mathcal{O}(\delta t^2)\bigg)(\cdot).
\end{align}
Now observe that
\begin{align}\label{eq:limit2}
    \lim_{\delta t\to 0} \mathscr{H}^{(u,v)}_n=(u\beta+v)\dot{\tilde{H}}_\lambda\big|_{\lambda=\lambda(t_n)}+u\dot{\beta} \ \tilde{H}_\lambda\big|_{\lambda=\lambda(t_n)}.
\end{align}
Combining~\eqref{eq:limit1} with~\eqref{eq:limit2} along with the time-spitting formula~\eqref{eq:split}, we can evaluate the continuum limit:
\begin{align}\label{eq:prop_before}
    \lim_{\delta t\to 0}\prod^0_{n=N}\mathcal{M}_u^{(n)}=\overleftarrow{\exp}\bigg(\int^\tau_0 dt \ \mathscr{L}_{\lambda} + \Upsilon^{(u,v)}_\lambda\star\bigg)(\cdot)=:\overleftarrow{\mathscr{P}}_{u,v}(\tau,0)(\cdot),
\end{align}
with
\begin{align}\label{eq:upsilon-1}
    \Upsilon^{(u,v)}_\lambda:=-\int_0^{(u\beta+v)/2} ds \  e^{-s\tilde{H}_\lambda} \dot{\tilde H}_\lambda e^{s \tilde{H}_\lambda}-\frac{u}{2}\dot{\beta} \ \tilde H_\lambda.
\end{align}
Now, let us note that
\begin{align}\label{eq:vanishing-trace-upsilon-pi}
\nonumber - \tr{\Upsilon^{(u,v)}_{\lambda} \star (\pi_{ \lambda}) }&=u\tr{(\beta\dot{\tilde{H}}_\lambda+\dot{\beta}\tilde{H}_\lambda)\pi_\lambda} + v\tr{\dot{\tilde{H}}_\lambda\pi_\lambda}, \\
\nonumber&=- u\tr{\dot{\pi}_\lambda} + v\tr{\dot{\tilde{H}}_\lambda\pi_\lambda}, \\
\nonumber &= v\tr{\dot{\tilde{H}}_\lambda\pi_\lambda} \\
&= v\tr{\dot{H}_\Lambda\pi_\lambda} - v \dot{F}_\lambda . 
\end{align} 
Taking the integral over time, therefore, yields
\begin{align}\label{eq:integral-current-eqm-vanish}
   - \int^\tau_0 dt \ \tr{\Upsilon^{(u,v)}_{\lambda(t)} \star (\pi_{ \lambda(t)}) }&=v\int^\tau_0 dt \ \tr{\dot{H}_{\Lambda(t)}\pi_{\lambda(t)}}-v\Delta F= v (\ww-\Delta F),
\end{align}
Now let us apply the definition of a shifted operator $\delta A_\lambda := A_\lambda -\tr{A_\lambda \pi_\lambda}\id$ to the current operator, defined in \eq{eq:upsilon-1}, to obtain 
\begin{align}\label{eq:Delta-upsilon-1}
\delta \Upsilon_{\lambda}^{(u,v)} := \Upsilon_{\lambda}^{(u,v)} - \tr{\Upsilon_{\lambda}^{(u,v)} \pi_{\lambda}}\id = -\int_0^{(u\beta+v)/2} ds \  e^{-s\tilde{H}_\lambda} \delta \dot{ H}_\lambda e^{s \tilde{H}_\lambda}-\frac{u}{2}\dot{\beta} \ \delta H_\lambda,
\end{align}
where we note that $\delta \tilde H_\lambda = \delta H_\lambda$ and $\delta \dot{\tilde H}_\lambda = \delta \dot H_\lambda$. It is then easy to show that \eqref{eq:integral-current-eqm-vanish} implies that the MGF obeys the identity
\begin{align}\label{eq:prop-after}
    \mathcal{G}_{\sigma,\tilde w}(u,v) &:= e^{v(\ww-\Delta F)}\tr{ \overleftarrow{\mathscr{P}}_{u,v}(\tau,0)(\pi_{\lambda_0})}, \nonumber \\
     &=   \tr{\exp \bigg(-\int^\tau_0 dt \ \Upsilon^{(u,v)}_{\lambda(t)} \star (\pi_{ \lambda(t)}) \bigg) \id \circ  \overleftarrow{\exp}\bigg(\int^\tau_0 dt \ \mathscr{L}_{\lambda} + \Upsilon^{(u,v)}_\lambda\star\bigg)(\pi_{\lambda_0})}, \nonumber \\
     & =  \tr{\overleftarrow{\exp}\bigg(\int^\tau_0 dt \ \mathscr{L}_{\lambda} + \delta \Upsilon^{(u,v)}_\lambda\star\bigg)(\pi_{\lambda_0})}, 
\end{align}
thus arriving at  \eqref{eq:MGFexact}.

\section{Derivation of~\eqref{eq:CGF2}}\label{app:b}

\

\noindent Using the perturbative Dyson series with the propagator in \eqref{eq:MGFexact},  we have
\begin{align}\label{eq:dyson2}
    \overleftarrow{\exp}\bigg(\int^\tau_0 dt \ \mathscr{L}_{\lambda} + \delta\Upsilon^{(u,v)}_{\lambda}\star\bigg)=\overleftarrow{P}^\tau_0+\sum^\infty_{n=1}\int_{0\leq t_1\leq ...\leq t_n\leq \tau}\overleftarrow{P}^\tau_{t_n}\circ\big(\delta\Upsilon^{(u,v)}_{\lambda(t_n)}\star\big)\overleftarrow{P}^{t_n}_{t_{n-1}}...\overleftarrow{P}^{t_2}_{t_1}\circ\big(\delta\Upsilon^{(u,v)}_{\lambda(t_1)}\star\big)\overleftarrow{P}^{t_1}_0,
\end{align}
where we define
\begin{align}
    \overleftarrow{P}^t_s:=\overleftarrow{\exp}\bigg(\int^t_s dt' \ \mathscr{L}_{\lambda(t')}\bigg).
\end{align}
 Since terms beyond $n=2$ in the sum will be at least order $\mathcal{O}(\epsilon^2)$, we are left with
\begin{align}\label{eq:dyson3}
    \nonumber\overleftarrow{\exp}\bigg(\int^\tau_0 dt \ \mathscr{L}_{\lambda} + \delta\Upsilon^{(u,v)}_{\lambda}\star\bigg)\simeq\overleftarrow{P}^\tau_0& + \int^\tau_0 dt_1 \ \overleftarrow{P}^\tau_{t_1}\circ\delta\Upsilon^{(u,v)}_{\lambda(t_1)}\star\overleftarrow{P}^{t_1}_0 \\
    &+\int^\tau_0 dt_2 \ \int^{t_2}_0 dt_1 \ \overleftarrow{P}^\tau_{t_2}\circ\delta\Upsilon^{(u,v)}_{\lambda(t_2)}\star\overleftarrow{P}^{t_2}_{t_1}\circ\delta\Upsilon^{(u)}_{\lambda(t_1)}\star\overleftarrow{P}^{t_1}_0.
\end{align}
Applying this propagator to the initial state $\rho_0 = \pi_{\lambda(0)}$, and taking the trace, we may write 
\begin{align}\label{eq:dyson4}
   \mathcal{G}_{\sigma,\tilde w}(u,v)\simeq 1+\mathcal{G}_1(u,v)+\mathcal{G}_2(u,v),
\end{align}
where we have defined
\begin{align}
    \label{eq:dyson5}&\mathcal{G}_1(u,v):=  \int^\tau_0 dt_1 \ \tr{\delta\Upsilon^{(u,v)}_{\lambda(t_1)}\star ( \rho_{t_1})}, \\
    \label{eq:dyson6}&\mathcal{G}_2(u,v):=  \int^\tau_0 dt_2 \ \int^{t_2}_0 dt_1 \ \tr{\delta\Upsilon^{(u,v)}_{\lambda(t_2)} \star \, \overleftarrow{P}^{t_2}_{t_1}\circ\delta\Upsilon^{(u,v)}_{\lambda(t_1)}\star(\rho_{t_1})}.
\end{align}
Here, we have used the fact that $\overleftarrow{P}^\tau_{t_n}$ is trace preserving. 
Let us first consider $\g_1(u,v)$ in the time coordinates $t' = \epsilon t_1$ which, by use of expansion~\eqref{eq:slow}  with  $dt_1 = dt'/ \epsilon$ and $\delta\Upsilon^{(u,v)}_{{\lambda}(t)} = \epsilon \delta\Upsilon^{(u,v)}_{\tilde{\lambda}(t')}$, can be written as 
\begin{align}\label{eq:g1-tprime}
  \nonumber\g_1(u,v) &=  \int^1_0 dt' \ \tr{\delta\Upsilon^{(u,v)}_{\tilde{\lambda}(t')} \star \big(\pi_{\tilde \lambda(t')}\big) } +   \epsilon\int^1_0 dt' \ \tr{\delta\Upsilon^{(u,v)}_{\tilde{\lambda}(t')}\star\mathscr{L}^+_{\tilde{\lambda}(t')}\big(\dot{\pi}_{\tilde{\lambda}(t')}\big) }+\mathcal{O}(\epsilon^2), \\
  &=\epsilon\int^1_0 dt' \ \tr{\delta\Upsilon^{(u,v)}_{\tilde{\lambda}(t')}\star\mathscr{L}^+_{\tilde{\lambda}(t')}\big(\dot{\pi}_{\tilde{\lambda}(t')}\big) }+\mathcal{O}(\epsilon^2).
\end{align}
From \eq{eq:Delta-upsilon-1} we observe that
\begin{align}\label{eq:current4}
    \delta\Upsilon_\lambda^{(u,v)}+(\delta\Upsilon_\lambda^{(u,v)})^\dagger &=  -\frac{(\beta u+v)}{2} \int_{-1}^1 ds \ e^{-(s (\beta u+v)/2)\tilde{H}_\lambda} \delta \dot{ H}_\lambda e^{(s (\beta u+v)/2)\tilde{H}_\lambda}-u\dot{\beta} \  \delta H_\lambda,
\end{align}
allowing us to, after converting back into the original time coordinates, reduce \eqref{eq:g1-tprime} to 
\begin{align}\label{eq:dyson7}
    \mathcal{G}_1(u,v)&\simeq \epsilon\int^1_0 dt' \ \tr{\delta\Upsilon^{(u,v)}_{\tilde{\lambda}(t')}\star\mathscr{L}^+_{\tilde{\lambda}(t')}(\dot{\pi}_{\tilde{\lambda}(t')}) }, \nonumber \\
    \nonumber&= - u \int^\tau_0 dt \ \dot{\beta}\tr{\delta H_{\lambda}\mathscr{L}^+_{\lambda}(\dot{\pi}_{\lambda}) } \\
    \nonumber& \ \ \ \ \ \ -\int^\tau_0 dt \ \frac{(\beta u+v)}{2}  \int_{-1}^1 ds \ \tr{e^{-(s (\beta u+v)/2)\tilde{H}_{\lambda}} \delta \dot{H}_{\lambda} e^{(s (\beta u+v)/2)\tilde{H}_{\lambda}}\mathscr{L}^+_{\lambda}(\dot{\pi}_{\lambda})}, \\
    \nonumber&= u \int^\tau_0 dt \int^\infty_0 d\theta \ \dot{\beta}\tr{e^{\theta\mathscr{L}_\lambda^*}(\delta H_{\lambda}) \ \dot{\pi}_{\lambda} } \\
    & \ \ \ \ \ \ +\int^\tau_0 dt \ \frac{(\beta u+v)}{2} \int^\infty_0 d\theta  \int_{-1}^1 ds \ \tr{e^{\theta\mathscr{L}_\lambda^*}\big(e^{-(s (\beta u+v)/2)\tilde{H}_{\lambda}} \delta \dot{H}_{\lambda} e^{(s (\beta u+v)/2)\tilde{H}_{\lambda}}\big) \ \dot{\pi}_{\lambda}}. 
\end{align}
Turning to the second contribution to the MGF~\eqref{eq:dyson6}, we first observe that the double integration can be separated according to
\begin{align}
    \mathcal{G}_2(u,v)=\mathcal{G}_2^{'}(u,v)+\mathcal{G}_2^{''}(u,v),
\end{align}
where
\begin{align}
   &\mathcal{G}_2^{'}(u,v)=\frac{1}{2}\int^\tau_0 dt_2 \ \int^{\tau}_0 dt_1 \ \tr{\delta\Upsilon^{(u,v)}_{\lambda(t_2)}\star\overleftarrow{P}^{t_2}_{t_1}\circ\delta\Upsilon^{(u)}_{\lambda(t_1)}\star(\rho_{t_1})}\Theta(t_2-t_1), \\
   &\mathcal{G}_2^{''}(u,v)=\frac{1}{2}\int^\tau_0 dt_1 \ \int^{\tau}_0 dt_2 \ \tr{\delta\Upsilon^{(u,v)}_{\lambda(t_1)}\star\overleftarrow{P}^{t_1}_{t_2}\circ\delta\Upsilon^{(u,v)}_{\lambda(t_2)}\star(\rho_{t_2})}\Theta(t_1-t_2),
\end{align} 
while $\Theta(x) = 0$ for $x<0$ and  $\Theta(x) =1$ for $x\geq 0$. Consider now the Taylor expansion of the Lindbladian around the point $t=t_1$:
\begin{align}
    \mathcal{L}_{\lambda(t)}=\mathcal{L}_{\lambda}\big|_{\lambda(t_1)}+\mathcal{O}(\dot{\lambda}(t_1)).
\end{align}
Substituting this into the time-ordered propagator in $\mathcal{G}_2^{'}(u,v)$ we find
\begin{align}
    \nonumber\mathcal{G}_2^{'}(u,v)&=\frac{1}{2}\int^\tau_0 dt_1 \ \int^{\tau}_0 dt_2 \ \bigg(\tr{\delta\Upsilon^{(u,v)}_{\lambda(t_2)}\star e^{(t_2-t_1)\mathscr{L}_{\lambda(t_1)}}\circ\delta\Upsilon^{(u,v)}_{\lambda(t_1)}\star(\rho_{t_1})}\Theta(t_2-t_1)+\mathcal{O}(\dot{\lambda}(t_1))\bigg), \\
    \nonumber&=\frac{1}{2}\int^\tau_0 dt_1 \ \int^{\tau-t_1}_{-t_1} d\theta \ \bigg(\tr{\delta\Upsilon^{(u,v)}_{\lambda(\theta+t_1)}\star e^{\theta\mathscr{L}_{\lambda(t_1)}}\circ\delta\Upsilon^{(u,v)}_{\lambda(t_1)}\star(\rho_{t_1})}\Theta(\theta)+\mathcal{O}(\dot{\lambda}(t_1))\bigg), \\
    \nonumber&=\frac{1}{2}\int^1_0 dt' \ \int^{\tau(1-t')}_{-\tau t'} d\theta \ \bigg(\tr{\delta\Upsilon^{(u,v)}_{\lambda(\theta+ t'/\epsilon)}\star e^{\theta\mathscr{L}_{\tilde{\lambda}( t')}}\circ\delta\Upsilon^{(u,v)}_{\tilde{\lambda}(t')}\star(\tilde{\rho}_{t'})}\Theta(\theta)+\mathcal{O}(\epsilon\dot{\tilde{\lambda}}(t_1))\bigg)\bigg), \\
    &=\frac{\epsilon}{2}\int^1_0 dt' \ \int^{\tau(1-t')}_{-\tau t'} d\theta \ \bigg(\tr{\delta\Upsilon^{(u,v)}_{\tilde{\lambda}(\epsilon\theta+ t')}\star e^{\theta\mathscr{L}_{\tilde{\lambda}( t')}}\circ\delta\Upsilon^{(u,v)}_{\tilde{\lambda}(t')}\star(\tilde{\rho}_{t'})}\Theta(\theta)+\mathcal{O}(\epsilon\dot{\tilde{\lambda}}(t_1))\bigg)\bigg),
\end{align}
where in the second line we introduced the variable $\theta=t_2-t_1$, and in the third line $t'=\epsilon t_1$. Note that $\lambda(\theta + t'/\epsilon) = \tilde{\lambda}(\epsilon \theta + t')$ and   $\dot{\lambda}(\theta+ t'/\epsilon) = \epsilon \dot{\tilde{\lambda}}(\epsilon \theta + t')$. Therefore, by taking the limit $\epsilon\to 0$ (ie. $\tau\to\infty$)  we have 
\begin{align}\label{eq:dyson8}
    \nonumber\mathcal{G}_2^{'}(u,v)&=\frac{\epsilon}{2}\int^1_0 dt' \ \int^{\infty}_{-\infty} d\theta \ \tr{\delta\Upsilon^{(u,v)}_{\tilde{\lambda}(t')}\star e^{\theta\mathscr{L}_{\tilde{\lambda}( t')}}\circ\delta\Upsilon^{(u,v)}_{\tilde{\lambda}(t')}\star(\tilde{\rho}_{t'})}\Theta(\theta)+\mathcal{O}(\epsilon^2), \\
    \nonumber&=\frac{\epsilon}{2}\int^1_0 dt' \ \int^{\infty}_{0} d\theta \ \tr{\delta\Upsilon^{(u,v)}_{\tilde{\lambda}(t')}\star e^{\theta\mathscr{L}_{\tilde{\lambda}( t')}}\circ\delta\Upsilon^{(u,v)}_{\tilde{\lambda}(t')}\star(\pi_{\tilde{\lambda}(t')})}+\mathcal{O}(\epsilon^2), \\
    &\simeq\frac{1}{2}\int^\tau_0 dt \ \int^{\infty}_{0} d\theta \ \tr{\delta\Upsilon^{(u,v)}_{\lambda}\star e^{\theta\mathscr{L}_{\lambda}}\circ\delta\Upsilon^{(u,v)}_{\lambda}\star(\pi_{\lambda})},
\end{align}
where in the second line we applied the slow driving expansion~\eqref{eq:slow}. By symmetry we also have $\mathcal{G}_2^{''}(u,v)=\mathcal{G}_2^{'}(u,v)$, so
\begin{align}
    \mathcal{G}_2(u,v) \simeq \int^\tau_0 dt \ \int^{\infty}_{0} d\theta \ \tr{\delta\Upsilon^{(u,v)}_{\lambda}\star e^{\theta\mathscr{L}_{\lambda}}\circ\delta\Upsilon^{(u,v)}_{\lambda}\star(\pi_{\lambda})}.
\end{align}
By substituting  \eq{eq:current4} into the first $\delta\Upsilon^{(u,v)}_{\lambda}\star$ in \eqref{eq:dyson8}, and then combining with \eqref{eq:dyson7}, we have
\begin{align}\label{eq:mgf_expand}
\mathcal{G}_{\sigma,\tilde w}(u,v) &\simeq 1 + u \int^\tau_0 dt\int^\infty_0 d\theta \ \dot{\beta}\tr{e^{\theta\mathscr{L}^*_\lambda}(\delta H_{\lambda})(\dot{\pi}_{\lambda}-\delta\Upsilon^{(u,v)}_{\lambda}\star(\pi_{\lambda})) } \\
    \nonumber& \ \ \ \ \ \ +\int^\tau_0 dt \ \frac{(\beta u+v)}{2} \int^\infty_0 d\theta  \int_{-1}^1 ds \ \tr{e^{\theta\mathscr{L}^*_\lambda}\big(e^{-(s (\beta u+v)/2)\tilde{H}_{\lambda}} \delta \dot{H}_{\lambda} e^{(s (\beta u+v)/2)\tilde{H}_{\lambda}}\big) \ \big(\dot{\pi}_{\lambda}-\delta\Upsilon^{(u,v)}_{\lambda}\star(\pi_{\lambda})\big)}.
\end{align}
Recall that, since we assume that $\lind_\lambda$ admits a privileged representation, and that the stationary state $\pi_\lambda$ is of Gibbs form, then $\lind_\lambda^*$ obeys time translation covariance \eq{eq:time-covariance}, $[\lind_\lambda^*, \mathscr{H}_\Lambda]=0$, where $\mathscr{H}_\Lambda (\cdot) := i [H_\Lambda, (\cdot)]$. Noting that for any $\alpha \in \re$,  $e^{-\alpha \tilde H_\lambda} (\cdot) e^{\alpha \tilde H_\lambda} = e^{\alpha i \mathscr{H}_\Lambda}(\cdot)$, time-translation covariance also implies that  $e^{\theta \lind_\lambda^*} \circ e^{\alpha i \mathscr{H}_\Lambda} = e^{\alpha i \mathscr{H}_\Lambda} \circ e^{\theta \lind_\lambda^*}$. Consequently, we may write
\begin{align}\label{eq:MGF_near}
\nonumber \mathcal{G}_{\sigma,\tilde w}(u,v)& \simeq 1- u \int^\tau_0 dt\int^\infty_0d\theta \ \dot{\beta}\tr{\delta H_{\lambda}(\theta)(\delta\Upsilon^{(u,v)}_{\lambda}\star(\pi_{\lambda})-\dot{\pi}_{\lambda}) } \\
    & \ \ \ \ \ \ -\int^\tau_0 dt \ \beta\int^\infty_0 d\theta  \int_{-\frac{( u+Tv)}{2}}^{\frac{( u+Tv)}{2}} ds \ \tr{ \delta \dot{H}_{\lambda}(\theta) \pi_\lambda^s \ \big(\delta\Upsilon^{(u,v)}_{\lambda}\star(\pi_{\lambda})-\dot{\pi}_{\lambda}\big)\pi_\lambda^{-s}}.
\end{align}
We next use the following identity:
\begin{align}
    \dot{\pi}_\lambda=-\dot{\beta}\delta H_\lambda \pi_\lambda-\beta\int^1_0 ds \ \pi^s_\lambda\delta \dot{H}_\lambda\pi^{1-s}_\lambda.
\end{align}
Furthermore, we also have
\begin{align}
    \delta\Upsilon^{(u,v)}_{\lambda}\star(\pi_{\lambda})=-\beta\bigg(\int^{\frac{( u+Tv)}{2}}_0+\int^{1}_{1-\frac{( u+Tv)}{2}}\bigg) dy  \ \pi_\lambda^y \  \delta \dot{H}_\lambda  \ \pi_\lambda^{1-y}-u\dot{\beta} \  \delta H_\lambda \pi_\lambda.
\end{align}
For the first term in~\eqref{eq:MGF_near} we therefore have
\begin{align}
\nonumber u \dot{\beta}\tr{\delta H_{\Lambda}(\theta)(\delta\Upsilon^{(u,v)}_{\lambda}\star(\pi_{\lambda})-\dot{\pi}_{\lambda}) }&=u(1-u)\dot{\beta}^2 \tr{\delta H_\lambda(\theta)\  \delta H_\lambda\pi_\lambda} \\
\nonumber& \ \ \ \ \ \ \ \ \ \ \ \ +u\dot{\beta}\beta\tr{\delta H_\lambda(\theta)\bigg(\int^1_0-\int^{\frac{( u+Tv)}{2}}_0-\int^{1}_{1-\frac{( u+Tv)}{2}}\bigg)dy \ \pi_\lambda^y \delta \dot{H}_\lambda \pi_\lambda^{1-y}}, \\
\nonumber&=u(1-u)\dot{\beta}^2 \tr{\delta H_\lambda(\theta)\ \delta H_\lambda\pi_\lambda} +u\dot{\beta}\beta\int^{1-\frac{( u+Tv)}{2}}_{\frac{( u+Tv)}{2}}dy \ \tr{\pi_\lambda^{-y}\delta H_\lambda(\theta) \pi_\lambda^y \delta \dot{H}_\lambda \pi_\lambda}, \\
\nonumber&=u(1-u)\dot{\beta}^2 \tr{\delta H_\lambda(\theta)\ \delta H_\lambda\pi_\lambda} +u\dot{\beta}\beta\int^{1-\frac{( u+Tv)}{2}}_{\frac{( u+Tv)}{2}}dy \ \tr{\delta H_\lambda(\theta)  \delta \dot{H}_\lambda \pi_\lambda}, \\
&=(u-u^2)\dot{\beta}^2 \tr{\delta H_\lambda(\theta)\ \delta H_\lambda\pi_\lambda} -\dot{\beta}\beta(u^2+Tv u -u) \tr{\delta H_\lambda(\theta)  \delta \dot{H}_\lambda \pi_\lambda},
\end{align}
where in the penultimate line we again used the commutation relation~\eqref{eq:time-covariance}. The second term in~\eqref{eq:MGF_near} can be evaluated as follows:
\begin{align}
\nonumber\beta\int_{-\frac{( u+Tv)}{2}}^{\frac{( u+Tv)}{2}} &ds \ \tr{ \delta \dot{H}_{\lambda}(\theta) \pi_\lambda^s \ \big(\delta\Upsilon^{(u,v)}_{\lambda}\star(\pi_{\lambda})-\dot{\pi}_{\lambda}\big)\pi_\lambda^{-s}}=(1-u)(u+Tv)\dot{\beta}\beta \tr{ \delta \dot{H}_{\lambda}(\theta)\delta H_\lambda \pi_\lambda} \\
\nonumber& \ \ \ \ \ \ \ \ \   +\beta^2\int_{-\frac{( u+Tv)}{2}}^{\frac{( u+Tv)}{2}} dx \int^{1-\frac{( u+Tv)}{2}}_{\frac{( u+Tv)}{2}}dy \ \tr{\delta \dot{H}_\lambda(\theta) \ \pi_\lambda^{y+x} \delta \dot{H}_\lambda \pi_\lambda^{1-x-y}}, \\
\nonumber&=(1-u)(u+Tv)\dot{\beta}\beta \tr{ \delta \dot{H}_{\lambda}(\theta)\delta H_\lambda \pi_\lambda} \\
\nonumber& \ \ \ \ \ \ \ \ \   +\beta^2\int_{0}^{ u+Tv} dx \int^{1-( u+Tv)}_{0}dy \ \tr{\delta \dot{H}_\lambda(\theta) \ \pi_\lambda^{y+x} \delta \dot{H}_\lambda \pi_\lambda^{1-(x+y)}}, \\
&=(1-u)(u+Tv)\dot{\beta}\beta \tr{ \delta \dot{H}_{\lambda}(\theta)\delta H_\lambda \pi_\lambda} \\
\nonumber& \ \ \ \ \ \ \ \ \   +\beta^2\int_{0}^{ u+Tv} ds \int^{1-s}_{s}ds' \ \tr{\delta \dot{H}_\lambda(\theta) \ \pi_\lambda^{s'} \delta \dot{H}_\lambda \pi_\lambda^{1-s'}},
\end{align}
where in the penultimate line we made the substitution $s'=x+y$ and $s=x$. Putting everything together in~\eqref{eq:MGF_near} leads to
\begin{align}\label{eq:MGF_near2} 
\nonumber\mathcal{G}_{\sigma,\tilde w}(u,v)&\simeq  1- \int^\tau_0 dt\bigg(  \beta^2  \bar{C}^{(u+Tv)}_\lambda(\dot{H}_\Lambda, \dot{H}_\Lambda) +(u-u^2) \dot{\beta}^2 C^{(0)}_\lambda( H_\Lambda,H_\Lambda)-(u^2+Tvu-u)\dot{\beta}\beta C^{(0)}_\lambda(H_\Lambda,\dot{H}_\Lambda) \\
&\ \ \ \ \ \ \ \ \ \ \ \ \ \ \ \ \ \ \ \ \ \ \ \ \ \ \ \ \ \ \ \ \ \ \ \ \ \ \ \ \ \ \ \ \  \ \ \ \ \ \ \ \ \ \ \ \ \ \  +(1-u)(u+Tv)\dot{\beta}\beta C^{(0)}_\lambda(\dot{H}_\Lambda,H_\Lambda)\bigg).
\end{align}
If we further assume that $\lind_\lambda$ satisfies the detailed balance condition~\eqref{eq:DB},   $\tilde{\mathscr{L}}_\lambda=\mathscr{L}_\lambda^*(\cdot) - 2 \mathscr{H}_\Lambda$  with    $\tilde{\mathscr{L}}_\lambda$ the s-dual generator given by~\eqref{eq:dual-reverse-gen}, then   we have the symmetry
\begin{align}\label{eq:DB_symm}
\nonumber C_\lambda^{(0)}(\dot{H}_\Lambda,H_\Lambda)&= \int_0^\infty d \theta \,  \tr{e^{\theta\mathscr{L}_\lambda^*}( \delta \dot{H}_\lambda)  \delta H_\lambda\pi_\lambda} =  \int_0^\infty d \theta \, \tr{e^{\theta\mathscr{L}_\lambda^*}( \delta \dot{H}_\lambda) \pi_\lambda  \delta H_\lambda}, \\
\nonumber&= \int_0^\infty d \theta \, \tr{ \delta\dot{H}_\lambda \pi_\lambda e^{\theta\tilde{\mathscr{L}}_\lambda}( \delta H_\lambda)} = \int_0^\infty d \theta \,  \tr{ \delta \dot{H}_\lambda \pi_\lambda e^{\theta(\mathscr{L}^*_\lambda -2 \mathscr{H}_\Lambda)}( \delta H_\lambda)}, \\
& = \int_0^\infty d \theta \,  \tr{ \delta \dot{H}_\lambda \pi_\lambda e^{\theta\mathscr{L}^*_\lambda} \circ e^{-2\theta  \mathscr{H}_\Lambda}( \delta H_\lambda)}, \nonumber \\
& = \int_0^\infty d \theta \,  \tr{ \delta \dot{H}_\lambda \pi_\lambda e^{\theta\mathscr{L}^*_\lambda} ( \delta H_\lambda)}, \nonumber \\
&=C_\lambda^{(0)}(H_\Lambda,\dot{H}_\Lambda),
\end{align}
where in the third line we have used time translation covariance \eq{eq:time-covariance}. Substituting this into~\eqref{eq:MGF_near2} and using $\ln{1+\epsilon}\simeq \epsilon$,  we arrive at the final expression for the CGF:
\begin{align}
    \mathcal{K}_{\sigma,\tilde w}(u,v)&\simeq  -\int_0^\tau dt \ \bigg(  \beta^2 \bar{C}^{(u+Tv)}_\lambda(\dot{H}_\lambda, \dot{H}_\lambda) +(u-u^2) \dot{\beta}^2 C^{(0)}_\lambda(H_\lambda,H_\lambda)+f_T(u,v)\dot{\beta}\beta \ C^{(0)}_\lambda(\dot{H}_\lambda, H_\lambda)\bigg).
\end{align}
As a consistency check, we note that throughout the derivation above each term in the perturbative expansion of order $\mathcal{O}(\epsilon^k)$ is also multiplied by contributions at least of the same order $\mathcal{O}((t^{eq})^k)$ in the relaxation time. This means we were able to drop all terms at least of order $\mathcal{O}(\epsilon^2)$ due to the slow driving assumption.

\section{Derivation of~\eqref{eq:evan}}\label{app:c}

\

Let us first observe that the integral fluctuation theorem implies the following:
\begin{align}
    \langle e^{-\sigma}\rangle=1 \implies \mathcal{K}_{\sigma,\tilde w}(1,0)=0.
\end{align}
Therefore we can infer from \eq{eq:CGF2} that
\begin{align}\label{eq:int1}
 \mathcal{K}_{\sigma,\tilde w}(1,0)  =   -\int^\tau_0 dt \  \beta^2 \bar{C}^{(1)}_\lambda(\dot{H}_\Lambda, \dot{H}_\Lambda) =0. 
\end{align}
We also see that
\begin{align}
f_T(u,v)=f_T(1-u,-v).
\end{align}
Expanding the CGF in~\eqref{eq:CGF2} then gives
\begin{align}
    \nonumber\mathcal{K}_{\sigma,\tilde w}(u,v)&=-\int^\tau_0 dt \ \int_0^{u+Tv} ds \int_s^{1-s} ds' \,  \mathcal{C}^{(s')}_\lambda(\dot{H}_\lambda, \dot{H}_\lambda)+(u-u^2) \dot{\beta}^2\mathcal{C}^{(0)}_\lambda(H_\lambda,H_\lambda)+f_T(u,v)\dot{\beta}\beta \ \mathcal{C}^{(0)}_\lambda(\dot{H}_\lambda,H_\lambda), \\
    \nonumber&=-\int^\tau_0 dt \ \int_1^{1-u-Tv} ds'' \int_{s''}^{1-s''} ds' \,  \mathcal{C}^{(s')}_\lambda(\dot{H}_\lambda, \dot{H}_\lambda)+(u-u^2) \dot{\beta}^2\mathcal{C}^{(0)}_\lambda(H_\lambda,H_\lambda)+f_T(u,v)\dot{\beta}\beta \ \mathcal{C}^{(0)}_\lambda(\dot{H}_\lambda,H_\lambda), \\
    \nonumber&=-\int^\tau_0 dt \ \bigg(\int_0^{1}+\int_1^{1-u-Tv}\bigg) ds'' \int_{s''}^{1-s''} ds' \,  \mathcal{C}^{(s')}_\lambda(\dot{H}_\lambda, \dot{H}_\lambda)+(u-u^2) \dot{\beta}^2\mathcal{C}^{(0)}_\lambda(H_\lambda,H_\lambda)+f_T(u,v)\dot{\beta}\beta \ \mathcal{C}^{(0)}_\lambda(\dot{H}_\lambda,H_\lambda), \\
    \nonumber&=-\int^\tau_0 dt \ \int_0^{1-u-Tv} ds'' \int_{s''}^{1-s''} ds' \,  \mathcal{C}^{(s')}_\lambda(\dot{H}_\lambda, \dot{H}_\lambda)+(u-u^2) \dot{\beta}^2\mathcal{C}^{(0)}_\lambda(H_\lambda,H_\lambda)+f_T(1-u,-v)\dot{\beta}\beta \ \mathcal{C}^{(0)}_\lambda(\dot{H}_\lambda,H_\lambda), \\
    &=\mathcal{K}_{\sigma,\tilde w}(1-u,-v),
\end{align}
where in the second line we made substitution $s''=1-s$, and in the third line we used~\eqref{eq:int1}. 

\section{Derivation of~\eqref{eq:CGF_ent}}\label{app:E}

\

Following the formalism given in Section~\ref{sec:2}, we assume the system obeys a Lindblad master equation~\eqref{eq:lind} with an invariant state $\mathscr{L}_\lambda(\pi_\lambda)=0$. We place no further assumption on the form of this state, and denote the corresponding non-equilibrium potential $\Phi_\lambda := -\ln{ \pi_\lambda}$. The only assumption we require is that the semi-group admits a privileged representation according to~\eqref{eq:priv2}. As shown in \cite{Fagnola2007}, this condition is weaker than the requirement of quantum detailed balance~\eqref{eq:DB} and may be applicable to systems with a non-thermal steady state. If one follows the same steps presented in Appendix~\ref{app:b} for the marginal distribution for the entropy production (namely setting $v=0$ in~\eqref{eq:mgf2}), we find the MGF to be
\begin{align}\label{eq:MGFexact_ent1}
    \mathcal{G}_{\sigma}(u)=\text{Tr}\bigg(\overleftarrow{\exp}\bigg(\int^\tau_0 dt \ \mathscr{L}_{\lambda} + \Upsilon^{(u)}_{\lambda}\star\bigg)(\pi_{\lambda(0)})\bigg),
\end{align}
where 
\begin{align}
    \Upsilon^{(u)}_\lambda:=-\int_0^{u/2} ds \  \pi^s_\lambda \ \dot{\Phi}_\lambda \pi_\lambda^{-s}.
\end{align}
From here we proceed to expand this expression up to first order in the driving speed, as done in Appendix~\ref{app:c}. Following similar steps that lead to~\eqref{eq:mgf_expand}, we can approximate~\eqref{eq:MGFexact_ent1} to yield  
\begin{align}\label{eq:mgf_expand2}
\mathcal{G}_{\sigma}(u) &\simeq 1  +\frac{u}{2} \int^\tau_0 dt \ \int^\infty_0 d\theta  \int_{-1}^1 ds \ \tr{e^{\theta\mathscr{L}^*_\lambda}\big(\pi_\lambda^{su/2} \dot{\Phi}_{\lambda} \pi_\lambda^{-su/2}\big) \ \big(\dot{\pi}_{\lambda}-\Upsilon^{(u)}_{\lambda}\star(\pi_{\lambda})\big)}.
\end{align}
We next consider the Fourier transform of the entropy production distribution, defined by
\begin{align}
    \tilde{\mathcal{G}}_\sigma(u):=\big\langle e^{i u \sigma} \big\rangle.
\end{align}
This is related to the MGF via a simple Wick rotation:
\begin{align}
    \tilde{\mathcal{G}}(u)=\mathcal{G}(-iu).
\end{align}
Applying this to~\eqref{eq:mgf_expand2} gives
\begin{align}\label{eq:mgf_expand3}
\tilde{\mathcal{G}}_{\sigma}(u) &= 1  -\frac{i u  }{2} \int^\tau_0 dt \ \int^\infty_0 d\theta  \int_{-1}^1 ds \ \tr{e^{\theta\mathscr{L}^*_\lambda}\big(\pi_\lambda^{-si u/2} \dot{\Phi}_{\lambda} \pi_\lambda^{siu/2}\big) \ \big(\dot{\pi}_{\lambda}-\Upsilon^{(-iu)}_{\lambda}\star(\pi_{\lambda})\big)}.
\end{align}
It will be useful to again introduce the automorphism on $\loh$ in~\eqref{eq:mod}, given by
\begin{align}
    \Omega_\lambda^{(t)}:=\pi_\lambda^{it}(.)\pi_\lambda^{-it}, \ \ t\in\mathbb{R}
\end{align}
As shown in \cite{Fagnola2007} (Lemma 3.2), assumption~\eqref{eq:mod} implies commutation with the generator $[\mathscr{L}^{*}_\lambda,\Omega_\lambda^{(t)}]=0 \ \forall t\in \mathbb{R}$, which also means
\begin{align}\label{eq:modcom}
    [e^{\theta\mathscr{L}_\lambda^{*}},\Omega_\lambda^{(t)}]=0.
\end{align}
We can apply this commutation relation to~\eqref{eq:mgf_expand3} and obtain the following:
\begin{align}
    \nonumber\tilde{\mathcal{G}}_{\sigma}(u)&=1  -\frac{i u  }{2} \int^\tau_0 dt \ \int^\infty_0 d\theta  \int_{-1}^1 ds \ \tr{e^{\theta\mathscr{L}^*_\lambda}\circ\Omega_\lambda^{(-su/2)}\big(\dot{\Phi}_{\lambda}\big) \ \big(\dot{\pi}_{\lambda}-\Upsilon^{(-iu)}_{\lambda}\star(\pi_{\lambda})\big)}, \\
    \nonumber &=1  -\frac{i u  }{2} \int^\tau_0 dt \ \int^\infty_0 d\theta  \int_{-1}^1 ds \ \tr{\Omega_\lambda^{(-su/2)}\circ e^{\theta\mathscr{L}^*_\lambda}\big(\dot{\Phi}_{\lambda}\big) \ \big(\dot{\pi}_{\lambda}-\Upsilon^{(-iu)}_{\lambda}\star(\pi_{\lambda})\big)}, \\
    &=1  -\frac{i u  }{2} \int^\tau_0 dt \ \int^\infty_0 d\theta  \int_{-1}^1 ds \ \tr{ e^{\theta\mathscr{L}^*_\lambda}\big(\dot{\Phi}_{\lambda}\big) \ \pi_\lambda^{-isu/2}\big(\dot{\pi}_{\lambda}-\Upsilon^{(-iu)}_{\lambda}\star(\pi_{\lambda})\big)\pi_\lambda^{isu/2}},
\end{align}
where we used cyclicity of the trace in the final line. Applying a Wick rotation again with $\mathcal{G}(u)=\tilde{\mathcal{G}}(iu)$, and a change of variables $x=su/2$ gives 
\begin{align}\label{eq:mgf_expand4}
    \mathcal{G}_{\sigma}(u) &= 1  + \int^\tau_0 dt \ \int^\infty_0 d\theta  \int_{-u/2}^{u/2} dx \ \tr{e^{\theta\mathscr{L}^*_\lambda}\big( \dot{\Phi}_{\lambda} \big) \ \pi^x_\lambda\big(\dot{\pi}_{\lambda}-\Upsilon^{(u)}_{\lambda}\star(\pi_{\lambda})\big)\pi^{-x}_\lambda}.
\end{align}
Next we can use the identities 
\begin{align}
    &\dot{\pi}_\lambda=-\int^1_0 dy \ \pi_\lambda^y \ \dot{\Phi}_\lambda \pi_\lambda^{1-y}, \\
    &\Upsilon^{(u)}_{\lambda}\star(\pi_{\lambda})=-\bigg(\int^{u/2}_0+\int^{1}_{1-u/2}\bigg) dy \ \pi_\lambda^y \ \dot{\Phi}_\lambda \pi_\lambda^{1-y},
\end{align}
which upon substitution into~\eqref{eq:mgf_expand4} leads to
\begin{align}
    \nonumber\mathcal{G}_{\sigma}(u) &= 1  - \int^\tau_0 dt \ \int^\infty_0 d\theta   \ \tr{e^{\theta\mathscr{L}^*_\lambda}\big( \dot{\Phi}_{\lambda} \big) \ \int_{-u/2}^{u/2} dx \bigg(\int^{u/2}_0 dy+\int^1_{1-u/2}dy-\int^1_0 dy\bigg) \ \pi_\lambda^{x+y} \ \dot{\Phi}_\lambda \pi_\lambda^{1-x-y}}, \\
    \nonumber&=1  +\int^\tau_0 dt \ \int^\infty_0 d\theta   \ \tr{e^{\theta\mathscr{L}^*_\lambda}\big( \dot{\Phi}_{\lambda} \big) \ \int_{-u/2}^{u/2} dx \int^{1-u/2}_{u/2} dy \ \pi_\lambda^{x+y} \ \dot{\Phi}_\lambda \pi_\lambda^{1-x-y}}, \\
    \nonumber&=1  +\int^\tau_0 dt \ \int^\infty_0 d\theta   \ \tr{e^{\theta\mathscr{L}^*_\lambda}\big( \dot{\Phi}_{\lambda} \big) \ \int_{0}^{u} dx \int^{1-u}_{0} dy \ \pi_\lambda^{x+y} \ \dot{\Phi}_\lambda \pi_\lambda^{1-x-y}}, \\
    &=1  +\int^\tau_0 dt \ \int^\infty_0 d\theta   \ \tr{e^{\theta\mathscr{L}^*_\lambda}\big( \dot{\Phi}_{\lambda} \big) \ \int_{0}^{u} ds \int^{1-s}_{s} ds' \ \pi_\lambda^{s'} \ \dot{\Phi}_\lambda \pi_\lambda^{1-s'}},
\end{align}
where we made the substitutions $s'=y+x$ and $s=x$ in the penultimate line. Finally, writing this in terms of the quantum covariance gives the final expression
\begin{align}
    \mathcal{G}_{\sigma}(u)=1-\int_0^\tau dt \   \bar{C}^{(u)}_\lambda(\dot{\Phi}_\lambda, \dot{\Phi}_\lambda),
\end{align}
Using $\ln{1+\epsilon}\simeq \epsilon$ completes the derivation.

\section{Single ion engine}\label{app:d}

\

For a fixed $\lambda=\{\beta, \omega\}$, the master equation for observables in the Heisenberg picture is given by the dual of \eq{eq:ion-lindblad}, which is
\begin{align}\label{eq:mastereq2}
    \mathscr{L}^{*}_\lambda(\cdot) &=i  \omega [a_\omega^\dagger a_\omega, (\cdot)]+\Gamma(N_\beta+1) \tilde{\mathcal{D}}_{a_\omega}[\cdot]+\Gamma \tilde{\mathcal{D}}_{a_\omega^{\dagger}}[\cdot],
\end{align}
with 
\begin{align}
    \tilde{\mathcal{D}}_X[\cdot]= X^{\dagger}(\cdot) X- \frac{1}{2}\{X^{\dagger}X,(\cdot)\}.
\end{align}
An observable $A$ evolved in the Heisenberg picture at a fixed control parameter $\lambda$ is thus given by \eq{eq:mastereq2} as $A(\theta)=e^{\theta \mathscr{L}^*_\lambda}(A)$.
Noting that $H_\omega = \omega(a_\omega^\dagger a_\omega + \frac{1}{2})$, we have $\tr{\dot{H}_\omega \pi_\lambda}= \dot{\omega} \partial_{\omega} F_{\lambda}$, where $F_{\lambda} = -\beta^{-1}\ln{\frac{e^{\beta \omega }}{e^{\beta \omega}-1}}$. As such, the adiabatic work is given by
\begin{align}
    \mathcal{W}=\int_0^\tau dt  \tr{\dot{H}_\omega \pi_\lambda}  = \int_0^\tau dt \, \dot{\omega}\,  \partial_{\omega} F_\lambda = \int_0^\tau dt \frac{\dot{\omega}}{e^{\beta \omega}-1}.
\end{align}

Using~\eqref{eq:CGF2}, we want to compute
\begin{align}
    \mathcal{K}_{\sigma,\tilde w}(u,v)&= -\int_0^\tau dt \ \bigg(  \beta^2 \bar{C}^{(u+Tv)}_\lambda(\dot{H}_\omega, \dot{H}_\omega) +(u-u^2) \dot{\beta}^2 C^{(0)}_\lambda(H_\omega,H_\omega)+f_T(u,v)\dot{\beta}\beta \ C^{(0)}_\lambda(\dot{H}_\omega, H_\omega)\bigg),
\end{align}
with 
\begin{align}
&C_\lambda^{(s)}(A,B):=\int^\infty_0 d\theta \ 
\text{cov}^{(s)}_\lambda\big(A(\theta),A(0)\big),
\nonumber\\
&    \text{cov}^{(s)}_\lambda(A,B):=\tr{A \ \pi^s_\lambda \  B \ \pi^{1-s}_\lambda}-\tr{A \ \pi_\lambda}\tr{B \ \pi_\lambda},
\end{align}
and:
\begin{align}
    \bar{C}_\lambda^{(y)}(A,B) :=\int_0^y ds \int_s^{1-s} ds' \ C_\lambda^{(s')}(A,B).
\end{align}
Note that
\begin{align}
    &x^2 = \frac{1}{2\omega}((a_\omega^{\dagger})^2+a_\omega^2+2a_\omega^{\dagger}a_\omega+1),
    \nonumber\\
    &p^2 = \frac{\omega}{2} \left( - (a_\omega^\dagger)^2 - a_\omega^2 +2a_\omega^\dagger a_\omega+1 \right),
    \nonumber\\
    &H_\omega = \omega\left( a_\omega^\dagger a_\omega +\frac{1}{2}\right).
\end{align}
We can then solve the master equation for each term individually, obtaining: 
\begin{align}
    &(a_\omega^{\dagger})^2(\theta)= e^{(2i\omega-\Gamma)\theta}(a_\omega^{\dagger})^2,
    \nonumber\\
    &a_\omega^2(\theta)= e^{(-2i\omega-\Gamma)\theta}a_\omega^2,
    \nonumber\\
    &a_\omega^{\dagger}a_\omega(\theta)=e^{-\Gamma \theta} a_\omega^{\dagger} a_\omega+N_\beta(1-e^{-\Gamma \theta}).
\end{align}
We then have:
\begin{align}
    &\int_0^{\infty} d\theta \hspace{1mm} x^2(\theta)=\frac{1}{2\omega}\left(\frac{(a_\omega^{\dagger})^2}{\Gamma-2i\omega}+\frac{a_\omega^2}{\Gamma+2i\omega}+2\frac{a_\omega^{\dagger}a_\omega}{\Gamma} \right)+c \id,
    \nonumber\\
   & \int_0^{\infty} d\theta \hspace{1mm} H(\theta)=\frac{a_\omega^{\dagger}a_\omega}{\Gamma} +c' \id,
\end{align}
with $c,c'$ constants. Terms proportional to $\id$ will disappear as $C_\lambda^{(y)}(\id,X)=0$ $\forall X$. A lengthy but straightforward calculation then yields:
\begin{align}
    &  {\rm cov}_\lambda^{(s)} ((a_\omega^{\dagger})^2,a_\omega^2)=\frac{2e^{2s\omega \beta}}{(e^{\beta \omega}-1)^2},
    \nonumber\\
       &  {\rm cov}_\lambda^{(s)} (a_\omega^2,(a_\omega^{\dagger})^2)={\rm cov}_\lambda^{(1-s)} ((a_\omega^{\dagger})^2,a_\omega^2),
        \nonumber\\
       & {\rm cov}_\lambda^{(s)} (a_\omega^\dagger a_\omega,a_\omega^\dagger a_\omega)=\frac{e^{\beta \omega}}{(e^{\beta \omega}-1)^2}.
\end{align}
Integrating them gives:
\begin{align}
    & \int_0^{y} dx \int_x^{1-x} ds\quad {\rm cov}_\lambda^{(s)} ((a_\omega^{\dagger})^2,a_\omega^2)=-\frac{2e^{\beta \omega}
   \sinh (\beta  (y-1) \omega) \sinh (\beta  y
   \omega)}{(e^{\beta \omega}-1)^2  \beta ^2 \omega^2},
    \nonumber\\
       & \int_0^{y} dx \int_x^{1-x} ds\quad {\rm cov}_\lambda^{(s)} (a_\omega^2,(a_\omega^{\dagger})^2)=\int_0^{y} dx \int_x^{1-x} dy\quad {\rm cov}_\lambda^{(s)} ((a_\omega^{\dagger})^2,a_\omega^2),
        \nonumber\\
       & \int_0^{y} dx \int_x^{1-x} ds\quad{\rm cov}_\lambda^{(s)} (a_\omega^\dagger a_\omega,a_\omega^\dagger a_\omega)=y(1-y)\frac{e^{\omega \beta}}{(e^{\omega\beta}-1)^2}.
\end{align}
Hence,
\begin{align}
&  \bar{C}^{(y)}_\lambda(\dot{H}_\omega, \dot{H}_\omega)  = \dot{\omega}^2 \frac{e^{\beta \omega}}{(e^{\beta \omega}-1)^2}  \left(-\frac{\Gamma \sinh{(\beta(y-1)\omega)} \sinh{(\beta y \omega)}}{\beta^2 \omega^2 (\Gamma^2+4\omega^2)}+\frac{y (1-y)}{\gamma} \right)  ,
 \nonumber\\
 &C^{(0)}_\lambda(H_\omega,H_\omega)=\frac{\omega^2}{\gamma} \frac{e^{\beta \omega}}{(e^{\beta \omega}-1)^2},
 \nonumber\\
 &C^{(0)}_\lambda(\dot{H}_\omega, H_\omega)=\frac{\omega \dot{\omega}}{\Gamma} \frac{e^{\beta \omega}}{(e^{\beta \omega}-1)^2}.
\end{align}
Putting everything together, we have
\begin{align}
    \mathcal{K}_{\sigma,\tilde w}(u,v) &= -\int_0^\tau dt \ \frac{e^{\beta \omega}}{(e^{\beta \omega}-1)^2} \nonumber \\
    & \qquad \qquad \times \bigg[  \beta^2\dot{\omega}^2   \left(-\frac{\Gamma \sinh{(\beta(u+\beta^{-1} v-1)\omega)} \sinh{(\beta (u+\beta^{-1}v) \omega)}}{\beta^2 \omega^2 (\Gamma^2+4\omega^2)}+\frac{1}{\Gamma}(u+\beta^{-1}v) (1-u-\beta^{-1}v) \right)
    \nonumber\\
    & \qquad \qquad \qquad  +(u-u^2) \dot{\beta}^2\frac{\omega^2}{\gamma}+(\beta^{-1}v-2u(u+\beta^{-1}v-1))\dot{\beta}\beta \ \frac{\omega \dot{\omega}}{\Gamma}\bigg],
\end{align}
from which we may obtain  the first two moments of work and  entropy production:
 \begin{align}
    \avg{w} &=   \ww + \int_0^\tau dt \ \frac{\dot{\omega} e^{\beta  \omega} \left(\omega \left(\Gamma ^2+4 \omega^2\right) (\dot{\beta}\omega+\beta 
   \dot{\omega})+\Gamma ^2 \dot{\omega} \sinh (\beta  \omega)\right)}{\Gamma  \omega \left(e^{\beta  \omega}-1\right)^2
   \left(\Gamma ^2+4 \omega^2\right)},
     \nonumber\\
  \avg{ \Delta w^2} &:= \avg{w^2} - \avg{w}^2 = \int_0^\tau dt \ \frac{2 \dot{\omega}^2 e^{\beta  \omega} \left(\Gamma ^2+4 \omega^2+\Gamma ^2 \cosh (\beta  \omega)\right)}{\left(e^{\beta 
   \omega}-1\right)^2 \left(\Gamma ^3+4 \Gamma  \omega^2\right)},
     \nonumber\\
     \avg{ \sigma} &=  \int_0^\tau dt \ \frac{e^{\beta  \omega} \left(\beta  \Gamma ^2 \dot{\omega}^2 \sinh (\beta  \omega)+\omega \left(\Gamma ^2+4 \omega^2\right)
   (\dot{\beta}\omega+\beta  \dot{\omega})^2\right)}{\Gamma  \omega \left(e^{\beta  \omega}-1\right)^2 \left(\Gamma
   ^2+4 \omega^2\right)},
     \nonumber\\
     \avg{\Delta \sigma^2} &:= \avg{\sigma^2} - \avg{\sigma}^2  = \int_0^\tau dt \ \frac{2 e^{\beta  \omega} \left(\omega \beta ^2 \Gamma ^2 \dot{\omega}^2 \cosh (\beta  \omega)+\omega \left(\Gamma ^2+4 \omega^2\right)
   (\dot{\beta}\omega+\beta  \dot{\omega})^2\right)}{\Gamma \omega \left(e^{\beta  \omega}-1\right)^2 \left(\Gamma^2+4 
   \omega^2\right)}.
 \end{align}
 
The FDR \eq{eq:FDR} is therefore given as
 \begin{align}
     2\Delta \mathcal{I}_\sigma &:= \avg{\Delta \sigma^2} - 2\avg{\sigma}, \nonumber \\
     & = \int_0^\tau dt \ \frac{ \beta \dot{\omega}^2 \Gamma (e^{2 \beta \omega} -1)\left(  \beta \omega \coth( \beta \omega )  - 1 \right)}{  \omega \left(e^{\beta  \omega}-1\right)^2 \left(\Gamma^2+4 
   \omega^2\right)}  \geqslant 0,
 \end{align}
with positivity guaranteed by the positivity of the integrand at all times $t$.

The quantum correction refining the efficiency bound in Ref.\cite{Miller2020a}, on the other hand, is given by the expression 
\begin{align}
    \Delta \mathcal{I}_w = \frac{1}{\tau}\int^\tau_0 dt \ \int^\infty_0 d\theta \  \tr{\delta \dot{H}_\lambda(\theta) \  \big(\mathbb{S}_\lambda-\mathbb{J}_\lambda\big)(\delta \dot{H}_\lambda(0))}.
\end{align}
Noting that the integrand can be rewritten in terms of covariances as
\begin{align}
\tr{A\big(\mathbb{S}_\lambda-\mathbb{J}_\lambda\big)(B)}=\frac{1}{2}\text{cov}^{(1)}_\lambda\big(A,B\big)+\frac{1}{2}\text{cov}^{(0)}_\lambda\big(A,B\big)-\int^1_0 ds \ \text{cov}^{(s)}_\lambda\big(A,B\big),
\end{align}
we thus obtain
\begin{align}
\Delta \mathcal{I}_w=\frac{1}{\tau}\int^\tau_0 dt \ \frac{\dot{\omega}^2 \Gamma (e^{2 \beta \omega} - 1)\left( \beta \omega \coth(\beta \omega) -1\right)}{2\beta \omega (e^{\beta \omega}-1)^2(\Gamma^2 + 4 \omega^2)}  \geqslant 0, 
\end{align} 
with positivity guaranteed by the positivity of the integrand for all $t$.

\end{document}